\documentclass[aps,pra,amsfonts, amssymb, amsmath,groupedaddress,showkeys,reprint,superscriptaddress,longbibliography,titleinbib]{revtex4-1}
\usepackage{orcidlink}
\usepackage[american]{babel} 
\usepackage{mathtools}
\usepackage{physics}
\usepackage{xfrac}
\usepackage{txfonts}
\usepackage{stmaryrd}
\usepackage[T1]{fontenc}
\usepackage{graphicx}
\usepackage{bm}
\usepackage{hyperref}

\usepackage{comment}
\definecolor{darkblue}{rgb}{0.,0.,0.}
\newcommand{\crr}[1]{{\color{darkblue}#1}}

\begin{document}
\title{Generating Entangled Steady States in Multistable Open Quantum Systems via Initial State Control}

\author{Diego Fallas Padilla}
\email{difa1788@colorado.edu}
\affiliation{JILA, NIST, and Department of Physics, University of Colorado, Boulder, CO, 80309, USA}
\affiliation{Center for Theory of Quantum Matter, University of Colorado, Boulder, CO 80309, USA}
\author{Raphael Kaubruegger}
\affiliation{JILA, NIST, and Department of Physics, University of Colorado, Boulder, CO, 80309, USA}
\affiliation{Center for Theory of Quantum Matter, University of Colorado, Boulder, CO 80309, USA}
\author{Adrianna Gillman}
\affiliation{Department of Applied Mathematics, University of Colorado, 526 UCB, Boulder, CO, 80309, USA}
\author{Stephen Becker}
\affiliation{Department of Applied Mathematics, University of Colorado, 526 UCB, Boulder, CO, 80309, USA}
\author{Ana Maria Rey}
\affiliation{JILA, NIST, and Department of Physics, University of Colorado, Boulder, CO, 80309, USA}
\affiliation{Center for Theory of Quantum Matter, University of Colorado, Boulder, CO 80309, USA}
\date{\today}

\begin{abstract}

Entanglement underpins the power of quantum technologies, yet it is fragile and typically destroyed by dissipation. Paradoxically, the same dissipation, when carefully engineered, can drive a system toward robust entangled steady states. However, this engineering task is nontrivial, as dissipative many-body systems are complex, particularly when they support multiple steady states. Here, we derive analytic expressions that predict how the steady state of a system evolving under a Lindblad equation depends on the initial state, without requiring integration of the dynamics. These results extend Refs.~\cite{Albert2014,Albert2016PRX}, showing that while the steady-state manifold is determined by the Liouvillian kernel, the weights within it depend on both the Liouvillian and the initial state. We identify a special class of Liouvillians for which the steady state depends only on the initial overlap with the kernel. Our framework provides analytical insight and a computationally efficient tool for predicting steady states in open quantum systems. As an application, we propose schemes to generate metrologically useful entangled steady states in spin ensembles via balanced collective decay.

\end{abstract}

\date{\today }
\maketitle

\section{Introduction} Dissipation is traditionally regarded as a major obstacle to quantum technologies, as it tends to destroy the fragile quantum coherence required for many applications. However, under carefully controlled conditions, dissipation can instead be harnessed as a resource. This idea underlies reservoir engineering, a paradigm in which the coupling between a system and its environment is tailored to actively drive the system toward a target steady state with desirable quantum properties such as entanglement~\cite{Poyatos1996,Rabl2004,Schirmer2010,Kienzler2015}.

Recent advances in experimental control have enabled the engineering of dissipation pathways that stabilize highly nontrivial quantum states—including entangled states~\cite{Hartmann2006,krauter2011}, logical encodings for quantum computation~\cite{verstraete2009quantum}, states with enhanced metrological sensitivity~\cite{Marzolino2014,Lu2015,Wang_2018,Mamaev2025,chu2025reconfigurable}, and even topologically ordered phases~\cite{Wang2024} from specific initial conditions. Fully exploiting these capabilities, however, requires more than numerical simulation: it calls for efficient and transparent methods to understand and predict the structure of steady states, particularly in many-body systems.

A key challenge stems from the fact that open quantum systems do not always exhibit a unique steady state. In these situations, engineering the system–environment coupling alone is insufficient; given a known Liouvillian superoperator—comprising both the Hamiltonian and the jump operators [see Eq.~\eqref{eq:1}]—it is equally important to understand how the steady state depends on the initial state preparation~\cite{Miranowicz2014}.

\begin{figure*}[th!]
\includegraphics[width=0.98\textwidth]{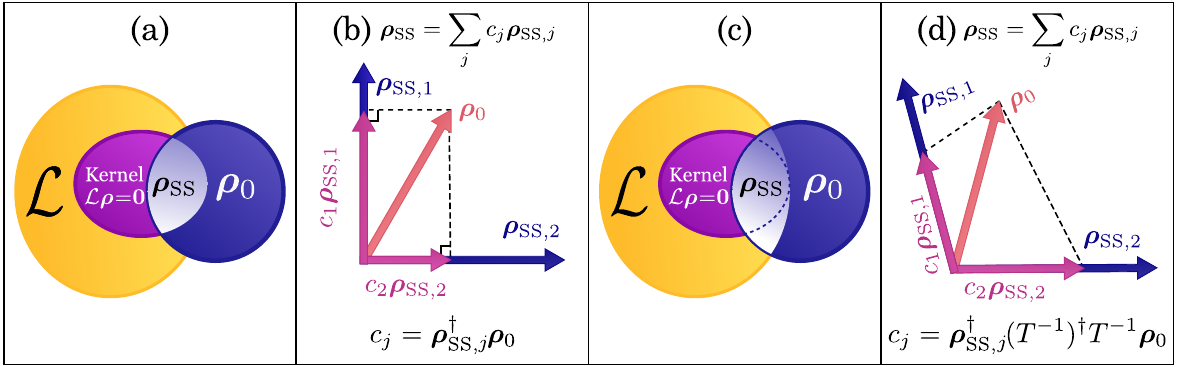}
\caption{\textbf{Role of initial states in multistable systems:} \textbf{(a)} In the case of a Hermitian Liouvillian $\mathcal{L}^{\dagger} = \mathcal{L}$, the steady state of the system $\bm{\rho}_{\rm SS}$ is completely determined by the overlap between the initial state $\bm{\rho}_0$ and the elements of the kernel of the Liouvillian [see Eq.~\eqref{result:specialcase}]. For this and panel (c), the yellow region corresponds to all the information contained in the Liouvillian. A smaller set of this area corresponds to the kernel, spanned by all vectors $\bm{\rho}$ that follow $\mathcal{L} \bm{\rho}=\bm{0}$, denoted by the purple region. The initial state is represented by the dark blue region, while the white region denotes the steady state. 
\textbf{(b)} For the case described in (a), the steady state is simply the sum of the projections of the initial state $\bm{\rho}_0$ into each of the vectors $\bm{\rho}_{{\rm SS},j}$ spanning the kernel. We note that in this case, since $T$ is unitary, the vectors $\bm{\rho}_{{\rm SS},j}$ can always be constructed to be orthogonal to each other. Moreover, in all examples studied here, these vectors correspond to valid density matrices, up to a constant factor, as seen in Eq.~\eqref{vectorsBalanced}, for example.
\textbf{(c)} For a general Liouvillian, the steady state is not fully determined by the overlap of the initial state and the elements of the kernel (enclosed with a dashed line), and additional information from the Liouvillian is required to predict the steady state. This information is encoded in the transformation $T$ for Eq.~\eqref{eq:result1} and in the elements of the kernel of $\mathcal{L}^{\dagger}$ ($\bm{\mathcal{U}}_j$) for Eq.~\eqref{eq:Albert}.
\textbf{(d)} For the more general case described in (c), the steady state is still a weighted sum of the vectors $\bm{\rho}_{{\rm SS},j}$ [Eq.~\eqref{eq:result1}], however, the weights $c_j$ are now defined by a different inner product $\bm{\rho}_{{\rm SS},j}^{\dagger} \mathcal{M}\, \bm{\rho}_0$, with $\mathcal{M}=(T^{-1})^\dagger T^{-1}$. We can think of $\mathcal{M}$ being a non-Euclidean metric that defines the inner product of vectors $\bm{\rho}_0$ and $\bm{\rho}_{{\rm SS},j}$. The non-Euclidean behavior is illustrated by the projections not being orthogonal (see dashed lines). For this case, the vectors $\bm{\rho}_{{\rm SS},j}$ are not necessarily orthogonal nor valid density matrices.}
\label{Fig1Sch}
\end{figure*}

In this work, we derive a set of analytic expressions, extending results from Refs.~\cite{Albert2014,Albert2016PRX}, that describe the steady state of a system evolving under a Lindblad equation for a given initial state. As depicted in Fig.~\ref{Fig1Sch}, these expressions share a common structure: the kernel of the Liouvillian superoperator determines a set of vectors spanning the steady-state manifold, while the weights of these vectors depend on both the initial state and additional Liouvillian properties not encoded in its kernel. We further identify a special class of Liouvillians for which the steady state can be predicted solely from the overlap of the initial state with the Liouvillian kernel [Fig.~\ref{Fig1Sch}(a)–(b)]. Understanding this structure not only deepens our perspective on dissipation as a resource, but also highlights initial-state preparation as a practical control knob for selecting steady states with robust entanglement or other properties relevant to quantum metrology and quantum computation.

Once the formalism is established, we apply it to the engineering of steady states in spin ensembles for quantum metrology~\cite{Pezze2018}. In particular, we build on the recent proposal of Ref.~\cite{Raphael2025}, which employs collective decay in a system of two spin ensembles to generate steady states suitable for differential sensing. We further show that introducing a second collective decay channel can enhance the differential phase sensitivity.

The remainder of the paper is organized as follows. In Sec.~\hyperref[SecII]{II}, we introduce the relevant formalism and discuss its applicability to different scenarios. In Sec.~\hyperref[SecIII]{III}, we illustrate the method with a simple example of two qubits evolving under collective jump operators. Building on the insights from this example and the recent work of Ref.~\cite{Raphael2025}, we propose in Sec.~\hyperref[SecIV]{IV} a scheme using collective jump operators to generate steady states in two ensembles of qubits, suitable for differential sensing. Importantly, in Secs.~\hyperref[SecIII]{III} and \hyperref[SecIV]{IV}, we benchmark the equivalence of the different expressions derived in Sec.~\hyperref[SecII]{II}. Finally, we summarize our findings in Sec.~\hyperref[SecV]{V}.

\section{Formalism} \label{SecII} 
The dynamics of a system in the presence of dissipation is often captured by the so-called Gorini-Kossakowski-Sudarshan-Lindblad (GKSL) master equation~\cite{lindblad1976generators,GKS1976}, which describes the time evolution of the density matrix of the system~\cite{breuer_Petruccione,rivas2012open,Manzano2020}. This equation applies under two key assumptions: (i) the system interacts weakly with its environment, and (ii) the environment acts as a Markovian bath, exhibiting no memory effects ($\hbar=1$):

\begin{eqnarray}\label{eq:1}
&\frac{d}{dt}\rho = -i [H,\rho] + \sum_{j} \gamma_j \left(L_j \rho L_j^{\dagger} - \frac{1}{2}(L_j^{\dagger}L_j\rho+\rho L_j^{\dagger}L_j)\right)\,, \nonumber \\
&\frac{d}{dt}\bm{\rho} = \mathcal{L}\,\bm{\rho}\,.
\end{eqnarray}

Here, $\rho$ is the density matrix of the system, $H$ is the Hamiltonian describing the coherent dynamics of the system (intra-system interactions), and $L_j$ are the different jump operators describing incoherent processes with rate $\gamma_j$ resulting from the interaction with the environment. $[A,B] = AB-BA$ denotes the commutator between operators $A$ and $B$, and $A^\dagger$ denotes the conjugate-transpose of $A$. In the second line, we have considered a vectorized form of this equation where $\bm{\rho}$ is the vectorized density matrix and $\mathcal{L}$ is the Liouvillian superoperator. Note that the Liouvillian comprises both the Hamiltonian and the jump operators as explicitly shown in Eq.~\eqref{liouvillianotimes}. Throughout the manuscript, bold symbols are used to denote vectors. This equation has been successfully used to study the properties of open quantum systems in quantum optics~\cite{carmichael2009open}, quantum computation~\cite{Lidar1998}, transport in biochemical systems~\cite{Plenio_2008,Moheni2008}, and condensed matter~\cite{prozen2011}.

To find the steady state(s) of the system, it is necessary to solve the algebraic equation $\frac{d}{dt} \bm{\rho} =\bm{0}$, which is equivalent to finding the eigenvectors with zero eigenvalue of the Liouvillian superoperator $\mathcal{L}$ (referred to as the kernel or null space of $\mathcal{L}$ \footnote{See Ref.~\cite{Campaioli2024} for nullspace numerical methods used for Liouvillians with multiple steady states.}). If the Liouvillian can be diagonalized with a transformation $T$ such that $T^{-1}\mathcal{L}T = \Lambda$, with $\Lambda$ being a diagonal matrix, the steady state of the system can be determined by (see \hyperref[app_A]{Appendix A} for derivation):

\begin{equation}
    \bm{\rho}_{\rm{SS}}  = \sum_{j=1}^n c_j \,\bm{\rho}_{{\rm SS},j},\, c_j = \bm{\rho}_{{\rm SS},j}^\dagger (T^{-1})^\dagger T^{-1} \bm{\rho}_0\,.
    \label{eq:result1}
\end{equation}

Here, $\bm{\rho}_{{\rm SS},j}$ correspond to the $n$ eigenvectors of $\mathcal{L}$ with zero eigenvalue. The weight $c_j$ of each of these eigenvectors in the steady state depends on their overlap with the initial state $\bm{\rho}_0$, where we note that an additional factor of $(T^{-1})^\dagger T^{-1} $ accounts for the transformation $T$ not necessarily being unitary. The eigenvectors $\bm{\rho}_{{\rm SS},j}$ are not required to be valid vectorized density matrices, moreover, they are not even required to be orthogonal to each other, namely, $\bm{\rho}_{{\rm SS},j}^{\dagger}\bm{\rho}_{{\rm SS},k}$ is not necessarily zero for $j \neq k$. The form of Eq.~\eqref{eq:result1} indicates that the steady state is a linear combination of the vectors $\bm{\rho}_{{\rm SS},j}$ that span the kernel of $\mathcal{L}$; however, to know the weights $c_j$ associated to each of these vectors we require more information than that contained in the kernel of $\mathcal{L}$ as evidenced by the $(T^{-1})^\dagger T^{-1} $ factor [see Fig.~\ref{Fig1Sch}(c)]. We can think of these weights as being defined by the inner product of $\bm{\rho}_0$ and $\bm{\rho}_{{\rm SS},j}$ in a space where the inner product is not equal to the dot product, but rather is defined through the matrix $\mathcal{M} = (T^{-1})^{\dagger} T^{-1}$. In that sense, we can think of $\mathcal{M}$ representing a non-Euclidean metric. This generalized projection of $\bm{\rho}_0$ into the $\bm{\rho}_{{\rm SS},j}$ vectors is illustrated in Fig.~\ref{Fig1Sch}(d).




Now, since the Liouvillian $\mathcal{L}$ might not be diagonalizable, a more general case is to consider its single value decomposition (SVD) in the form $\mathcal{L} = U \Sigma V^{\dagger}$, where $\Sigma$ is a diagonal matrix and both $U$ and $V$ are unitary matrices. In this case, we can express the steady state of the system through the equation (see \hyperref[app_A]{Appendix A} for derivation):

\begin{equation}
\bm{\rho}_{\rm SS} = \sum_{j=1}^n \tilde{c}_j \bm{\mathcal{V}}_{j},\, \tilde{c}_j = \bm{\mathcal{U}}_j^{\dagger}  \bm{\rho}_0\,.
\label{eq:Albert}
\end{equation}

Here, $\mathcal{L}$ has $n$ singular values equal to zero, $\bm{\mathcal{U}}_j$ and $\bm{\mathcal{V}}_j$ are the column vectors of $U$ and $V$ corresponding to a zero singular value, respectively. Moreover, these sets are constructed to fulfill the biorthogonality condition $\bm{\mathcal{U}}_j^{\dagger} \bm{\mathcal{V}}_k = \delta_{jk} $. In \hyperref[app_B]{Appendix B}, we provide a step-by-step explanation of how to obtain these biorthogonal sets for any $\mathcal{L}$.

Eq.~\eqref{eq:Albert} was first reported in Refs.~\cite{Albert2014, Albert2016PRX} in the context of conserved quantities. Each $\bm{\mathcal{U}}_j$ represents a conserved quantity as they fulfill the equation $\mathcal{L}^{\dagger} \bm{\mathcal{U}}_j = 0$. Consequently, a system with more than one steady state (with more than one singular value equal to zero) necessarily has at least one conserved quantity different from the identity. The interpretation of Eq.~\eqref{eq:Albert} in this context is that since all $\bm{\mathcal{U}}_{j}$ are conserved, we can use the initial values of these conserved quantities ($\tilde{c}_j$) to predict the steady state of the system. The same structure discussed for Eq.~\eqref{eq:result1} repeats here: Although the kernel of $\mathcal{L}$ contains all the information about the vectors $\bm{\mathcal{V}}_j$, additional information is required to infer the weights $\tilde{c}_j$, which define the linear superposition of states spanning the steady state. The extra information is contained in the kernel of $\mathcal{L}^{\dagger}$. 

We should note that a special case occurs when the Liouvillian is Hermitian, i.e., $\mathcal{L} = \mathcal{L}^{\dagger}$ [see Fig.~\ref{Fig1Sch}(a)]. In that case, the transformation $T$ is necessarily unitary, and Eq.~\eqref{eq:result1} reduces to:
\begin{equation}
\bm{\rho}_{\rm{SS}}  = \sum_{j=1}^n \bm{\rho}_{{\rm SS},j}^\dagger \bm{\rho}_0\, \bm{\rho}_{{\rm SS},j}
\label{result:specialcase}
\end{equation}which is a conventional vector projection of the initial state into the vectors $\bm{\rho}_{{\rm SS},j}$ as shown in Fig.~\ref{Fig1Sch}(b). In this case, all the information required to predict the steady state of the system is contained in the kernel of $\mathcal{L}$. Since $T$ is unitary, the vectors $\bm{\rho}_{{\rm SS},j}$ can always be constructed to be orthogonal to each other. As we show later, in all examples studied here, the vectors $\bm{\rho}_{{\rm SS},j}$ correspond to valid density matrices, up to a normalization constant (see Eqs.~\eqref{vectorsBalanced} and~\eqref{balancedDicke}, for example). An example of a Hermitian Liouvillian occurs when $H=0$ and there are two jump operators such that $L_1=L_2^{\dagger}$ and $\gamma_1=\gamma_2$ (see Eq.~\eqref{liouvillianotimes} in \hyperref[app_B]{Appendix B}). Having $H \neq 0$ will generally lead to a non-Hermitian Liouvillian; however, $\mathcal{L}$ might still be complex symmetric under certain conditions, and an expression equivalent to Eq.~\eqref{result:specialcase} can be found in that case \footnote{If $H$, $L_1$ and $L_2$ are real, with $L_1=L_2^{\dagger}$ and $\gamma_1=\gamma_2$, the Liouvillian is a complex symmetric matrix that can be diagonalized with a Takagi factorization $\mathcal{L} = S \Lambda S^T$, where $S$ is unitary and $\Lambda$ is diagonal. Note that the Takagi factorization is a special case of the SVD where $U = V^*$. An expression equivalent to Eq.~\eqref{result:specialcase} can be found in this case following a similar procedure as in \hyperref[app_A]{Appendix A}.}.

The form of Eqs.~\eqref{eq:result1},~\eqref{eq:Albert}, and~\eqref{result:specialcase} allows us to draw analogies and distinctions with multistable classical systems. In nonlinear classical systems governed by a differential equation $\frac{d \bm{x}}{dt} = f(\bm{x})$, solving $f(\bm{x}_{\rm F}) = 0$ yields the fixed points $\bm{x}_{\rm F}$. If these points are stable attractors, the system asymptotically approaches one of them at long times. The set of initial states that converge to a given attractor defines its basin of attraction, whose boundaries generally depend in a highly nontrivial way on all system parameters, making their characterization particularly challenging~\cite{Grebogi1986,eschenazi1989,Kozinsky2007}.

In the quantum case, analogous structures emerge: the vectors $\bm{\mathcal{V}}_j$ and $\bm{\rho}_{{\rm SS},j}$ spanning the kernel of $\mathcal{L}$ determine the basis vectors of the steady state, with weights that are, in general, complicated functions of the system parameters (encoded in $T$ or $\bm{\mathcal{U}}_j$). A key difference from the classical case is that the coefficients $c_j$ and $\tilde{c}_j$ determine the contribution of each vector to the final state, rather than deterministically selecting a single one. For some solutions of the form of Eq.~\eqref{result:specialcase}, where all $\bm{\rho}_{{\rm SS},j}$ can correspond to valid density matrices, these vectors might be interpreted as attractors \crr{as we discuss more in depth in \hyperref[app_C]{Appendix C.}} For those cases, each individual quantum trajectory converges to exactly one attractor $\bm{\rho}_{{\rm SS},j}$, with the probability of ending in that attractor given by $c_j$. The final state of the system is thus a statistical mixture over many such trajectories, illustrating the inherently probabilistic nature of quantum dynamics. \crr{More generally, this notion of attractors is not applicable to all quantum multistable systems as we exemplify in \hyperref[app_C]{Appendix C.}.}

Before illustrating how Eqs.~\eqref{eq:result1},~\eqref{eq:Albert}, and~\eqref{result:specialcase} can be used to build intuition for optimizing a system’s initial state to achieve a steady state with desirable properties, we present an additional expression that predicts the steady state generated by an arbitrary Liouvillian (see \hyperref[app_A]{Appendix A} for the derivation):

\begin{equation}
\bm{\rho}_{\rm SS} = \left(\lim_{s\rightarrow 0} s (s\mathbb{I} - \mathcal{L})^{-1}\right)\bm{\rho}_0 \approx  \left(\mathbb{I} - \frac{1}{\epsilon}\mathcal{L}\right)^{-1}\bm{\rho}(0 )\,,
\label{eq:result2}
\end{equation}
where $\mathbb{I}$ is the identity matrix. In the final expression, we have approximated the real variable $s$ by a numerical parameter $\epsilon \ll 1$. In that sense, $\epsilon$ controls how good our approximation is, if the right-hand side is solved numerically. The form of this expression is very similar to shift-invert techniques~\cite{Luitz2015,Pietracaprina2018} used to diagonalize many-body Hamiltonians, which highlights that the determination of the steady state relies heavily on the zero-eigenvalue eigenvectors (kernel) of the Liouvillian ($\epsilon \ll 1$), in agreement with the interpretation of the previous methods discussed here.

As we exemplify later, Eqs.~\eqref{eq:result1},~\eqref{eq:Albert}, and~\eqref{result:specialcase} might allow us to obtain closed-form expressions for the steady states in certain cases. From a numerical perspective, the slowest step in computing these equations corresponds to computing and storing $T$, $\bm{\mathcal{U}}_j$, or $\bm{\mathcal{V}}_j$. Once this step is complete, the evaluation of many initial conditions $\bm{\rho}_0$ is straightforward. The same holds for Eq.~\eqref{eq:result2} if we pre-compute and store the inverse of the matrix $A = \mathbb{I} - \tfrac{1}{\epsilon}\mathcal{L}$. In this approach, deciding which equation to use reduces to identifying which of these bottleneck processes is faster or most memory efficient. Alternatively, depending on the structure of $A$, it may be more numerically stable and efficient to compute $A^{-1}\bm{\rho}_0$ directly using either direct (some Gaussian elimination variant) or iterative methods. This approach can be significantly faster if the number of initial states to evaluate is not too large. In all the examples below, we numerically compute the steady state using Eq.~\eqref{eq:result2} and show that this expression is equivalent to all other equations presented above. The numerical utility of this expression is further discussed in the Supplemental Material~\cite{supplemental}.

\section{Two qubits with collective dissipation} \label{SecIII}
\begin{figure}[t!]
\includegraphics[width=0.46\textwidth]{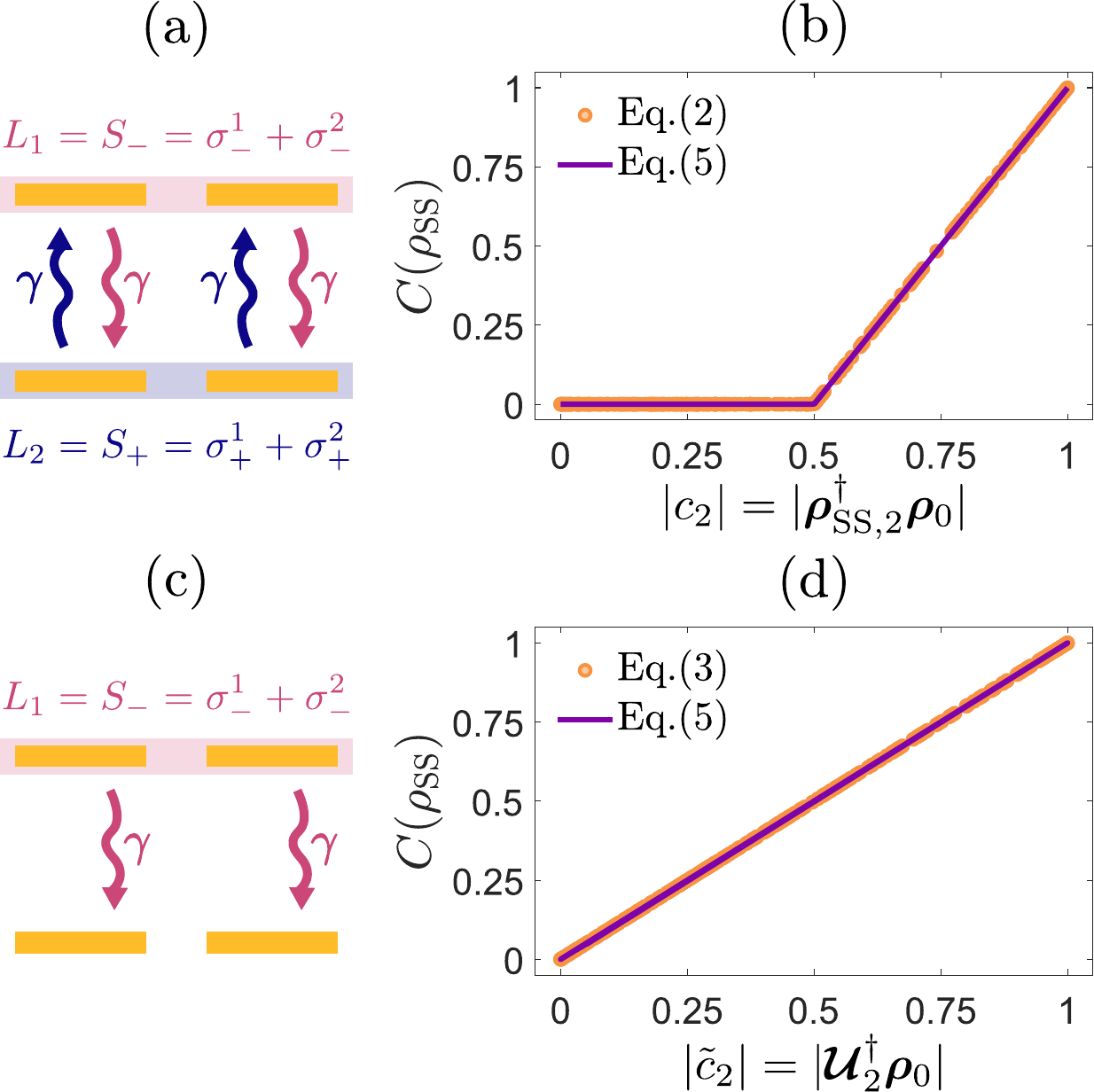}
\caption{\textbf{Two qubits under collective dissipation:} \textbf{(a)} Two qubits experience collective dissipation through jump operators $L_1=S_-=\sigma_-^1+\sigma_-^2$ and $L_2 =S_+ =\sigma_+^1+\sigma_+^2$. The two processes have the same rates $\gamma_1=\gamma_2=\gamma$. \textbf{(b)} Concurrence in the steady state as a function of the overlap $c_2 = \bm{\rho}_{{\rm SS},2}^\dagger \bm{\rho_0}$ for the setup in (a). Data represents 300 initial states $\bm{\rho}_0$ sampled at random. The steady state is computed using Eq.~\eqref{eq:result2} with $\epsilon = 0.0001$ (solid line), and Eq.~\eqref{eq:result1} (markers). \textbf{(c)} Same as in (a), but the qubits only undergo collective decay with jump operator $L_1$.  \textbf{(d)} Concurrence in the steady state as a function of the overlap $\tilde{c}_2 = \bm{\mathcal{U}}_2^{\dagger}\bm{\rho}_0$ for the setup in (c). Data represents 300 initial states $\bm{\rho}_0$ sampled at random. The steady state is computed using Eq.~\eqref{eq:result2} with $\epsilon = 0.0001$ (solid line), and Eq.~\eqref{eq:Albert} (markers).}
\label{Fig2q}
\end{figure}

In this section, we consider two examples to illustrate how the expressions described above can be utilized. First, we consider a case where the condition $\mathcal{L}=\mathcal{L}^{\dagger}$ is met, making it suitable to use Eq.~\eqref{result:specialcase}. The system is composed of two qubits, each one consisting of two states labeled as $\ket{0}$ and $\ket{1}$. We describe the full Hilbert space of the two qubits in the product basis $\{\ket{1} = \ket{1}\otimes \ket{1}, \ket{2}=\ket{1} \otimes \ket{0}, \ket{3} = \ket{0}\otimes \ket{1}, \ket{4}=\ket{0}\otimes \ket{0}\}$.  The system evolves under the action of two collective jump operators $L_1 = S_-=\sigma_-^1+\sigma_-^2$ and $L_2 = S_+=\sigma_+^1+\sigma_+^2$ with the same rate $\gamma_1=\gamma_2=\gamma$ [see Eq.\eqref{eq:1}]. Here, $\sigma_-^j = (\sigma_x^j-i\sigma_y^j)/2$ and $\sigma_+^j = (\sigma_x^j+i \sigma_y^j)/2$ are defined in terms of the Pauli matrices in the $x$- and $y$-directions acting on the $j$-th qubit. A schematic representation of the system is shown in Fig.~\ref{Fig2q}(a). The simultaneous presence of collective lowering and raising jump operators can be observed, for example, in the study of superradiance in cavity QED systems with two dressing tones~\cite{luo2024realization,luo2025hamiltonian}. Diagonalizing $\mathcal{L}$ one finds the vectors:

\begin{eqnarray}
&&\rho_{{\rm SS},1} = -\sqrt{3}\left(\frac{1}{3}\ket{1}\bra{1}+\frac{1}{3}\ket{4}\bra{4}+\frac{1}{3}\ket{\phi_+}\bra{\phi_+}\right), \nonumber \\
&&\rho_{{\rm SS},2} = \ket{\phi_-}\bra{\phi_-},\quad \ket{\phi_\pm} = \frac{1}{\sqrt{2}} (\ket{2}\pm \ket{3}).
\label{vectorsBalanced}
\end{eqnarray}

For simplicity we use a matrix representation of $\bm{\rho}_{{\rm SS},j}$ for all examples. In this case, $\bm{\rho}_{{\rm SS},2}$ represents a valid density matrix (the single state $\ket{\phi_-}$), while $\bm{\rho}_{{\rm SS},1}$ can be turned into a valid density matrix by rescaling it with a constant factor $-1/\sqrt{3}$. Clearly, this would cause $(T^{-1})^\dagger T^{-1}$ to not be equal to the identity but rather a diagonal matrix. In this case, all the information is contained in the vectors $\bm{\rho}_{{\rm SS},j}$; consequently, the symmetries of the system are explicitly visible in them. For instance, both jump operators commute with the total spin length, namely, $[S^2,S_-]=[S^2,S_+]=0$, where $S^2 = S_x^2+S_y^2+S_z^2$ and $S_{\alpha} = (\sigma_\alpha^1 + \sigma_{\alpha}^2)/2$, for $\alpha=x,y,z$. This means that the dynamics do not mix different spin sectors. This is shown in Eq.~\eqref{vectorsBalanced}, where $\bm{\rho}_{{\rm SS},2}$ represents the state with eigenvalue $S=0$, while $\bm{\rho}_{{\rm SS},1}$ is an equal mixture of all states with $S=1$. We also note that $\langle S_z \rangle_{\rm SS} ={\rm tr}((\sigma_z^1+\sigma_z^2)/2\,  \rho_{\rm SS})=0$, reflecting the balanced character of the two jump operators $L_1$ and $L_2$

In this case, the optimization of the initial state is straightforward if we want to maximize the entanglement of the steady state. $\bm{\rho}_{{\rm SS},2}$ is a maximally entangled state, while $\bm{\rho}_{{\rm SS},1}$ has reduced entanglement due to being a mixture. Then, to maximize the entanglement, we need to maximize the overlap $c_2=\bm{\rho}_{{\rm SS},2}^\dagger\bm{\rho}_0$. To confirm this, we sample random initial states $\bm{\rho}_0$ and compute the entanglement in the generated steady state as a function of the overlap, as shown in Fig.~\ref{Fig2q}(b). To quantify the entanglement of the steady state, we use the concurrence, mathematically defined by~\cite{Hill1997,Wootters1998}:

\begin{equation}
    C(\rho_{\rm SS}) = \text{max}(0, \mu_1 - \mu_2 - \mu_3 - \mu_4)\,.
    \label{cur}
\end{equation}

Here $\rho_{\rm SS}$ is the steady-state density matrix and $\mu_j$ correspond to the eigenvalues of  $R = \sqrt{\sqrt{\rho_{\rm SS}}\tilde{\rho}\sqrt{\rho_{\rm SS}}}$, where $\tilde{\rho}= S \rho_{\rm SS}^* S$ and  $S = \sigma_y \otimes \sigma_y$. $\sigma_y$ is the Pauli matrix in the $y$-direction. The eigenvalues are defined in descending order, namely, $\mu_1 \geq \mu_2 \geq \mu_3 \geq \mu_4$. If $C(\rho_{\rm SS})=0$ the state is unentangled, and if $C(\rho_{\rm SS})=1$ the state is maximally entangled. As expected, the entanglement in the system increases as the overlap with the singlet state $\bm{\rho}_{{\rm SS},2}^\dagger \bm{\rho_0}$ becomes larger. Due to the form of $\bm{\rho}_{\rm SS}$ [see Eq.~\eqref{result:specialcase}] corresponding to the vectors in Eq.~\eqref{vectorsBalanced}, three of the eigenvalues $\lambda_j$ in Eq.~\eqref{cur} are identical, then, for $C(\rho_{\rm SS}) >0$ we require $\lambda_1 > 0.5$ which implies $\bm{\rho}_{{\rm SS},2}^\dagger \bm{\rho_0}>0.5$. Once this condition is met, the entanglement in the steady state grows linearly with the overlap.

Now, let's consider that $\gamma_2=0$, meaning that we only have a single jump operator $L_1 = S_-=\sigma_-^1+\sigma_-^2$ as depicted in Fig.~\eqref{Fig2q}(c). In this case, Eq.~\eqref{result:specialcase} is not valid anymore. Moreover, $\mathcal{L}$ is not diagonalizable, hence, we use Eq.~\eqref{eq:Albert} for this case. The explicit form of the vectors $\bm{\mathcal{V}}_j$ is found to be:

\begin{eqnarray}
&&\mathcal{V}_1 =\ket{4}\bra{4},\quad \mathcal{V}_2 = \ket{\phi_-}\bra{\phi_-},\nonumber \\
&&\quad \mathcal{V}_3 = \mathcal{V}_4^\dagger =  \ket{4}\bra{\phi_-}.
\end{eqnarray}

In this case, $\bm{\mathcal{V}}_1$ and $\bm{\mathcal{V}}_2$ both represent two valid density matrices corresponding to the dark states of $L_1$, namely, $L_1\ket{4} = L_1\ket{\phi_-}=0$, while $\bm{\mathcal{V}}_3$ and $\bm{\mathcal{V}}_4$ are the coherences between these two states. In this case, the weights $\tilde{c}_j$ are given by the overlaps between $\bm{\mathcal{U}}_j$ and $\bm{\rho}_0$ for all $j$ [Eq.~\eqref{eq:Albert}].  We can choose $\bm{\mathcal{V}}_j = \bm{\mathcal{U}}_j$ for $j=2,3,4$, if $\mathcal{U}_1 = \ket{1}\bra{1}+\ket{\phi_+}\bra{\phi_+}+\ket{4}\bra{4}$. We note that $\bm{\mathcal{U}}_1 \neq \bm{\mathcal{V}}_1$. If the system is initialized in the state $\rho_0 = \ket{1}\bra{1}$, the steady state is $\bm{\mathcal{V}}_1$ since $[L_1,S^2]=0$. However, the overlap $\bm{\mathcal{U}}_1^\dagger \bm{\rho}_0$ vanishes, confirming that $\bm{\mathcal{U}}_1 \neq \bm{\mathcal{V}}_1$. Since $\bm{\mathcal{V}}_2$ represents a maximally entangled state and $\bm{\mathcal{V}}_1$ a separable state (no entanglement), we expect the entanglement of the steady state to grow with the overlap $\bm{\mathcal{V}}_2^\dagger \bm{\rho}_0$ since $\bm{\mathcal{V}}_2 = \bm{\mathcal{U}}_2$. This is numerically confirmed in Fig.~\ref{Fig2q}(d), using the concurrence in randomly sampled states. We note that in Fig.~\ref{Fig2q} panels (b) and (d), the results obtained using Eq.~\eqref{eq:result2} are consistent with those obtained through Eqs.~\eqref{eq:result1}-\eqref{result:specialcase}.

\begin{figure*}[th!]
\includegraphics[width=0.98\textwidth]{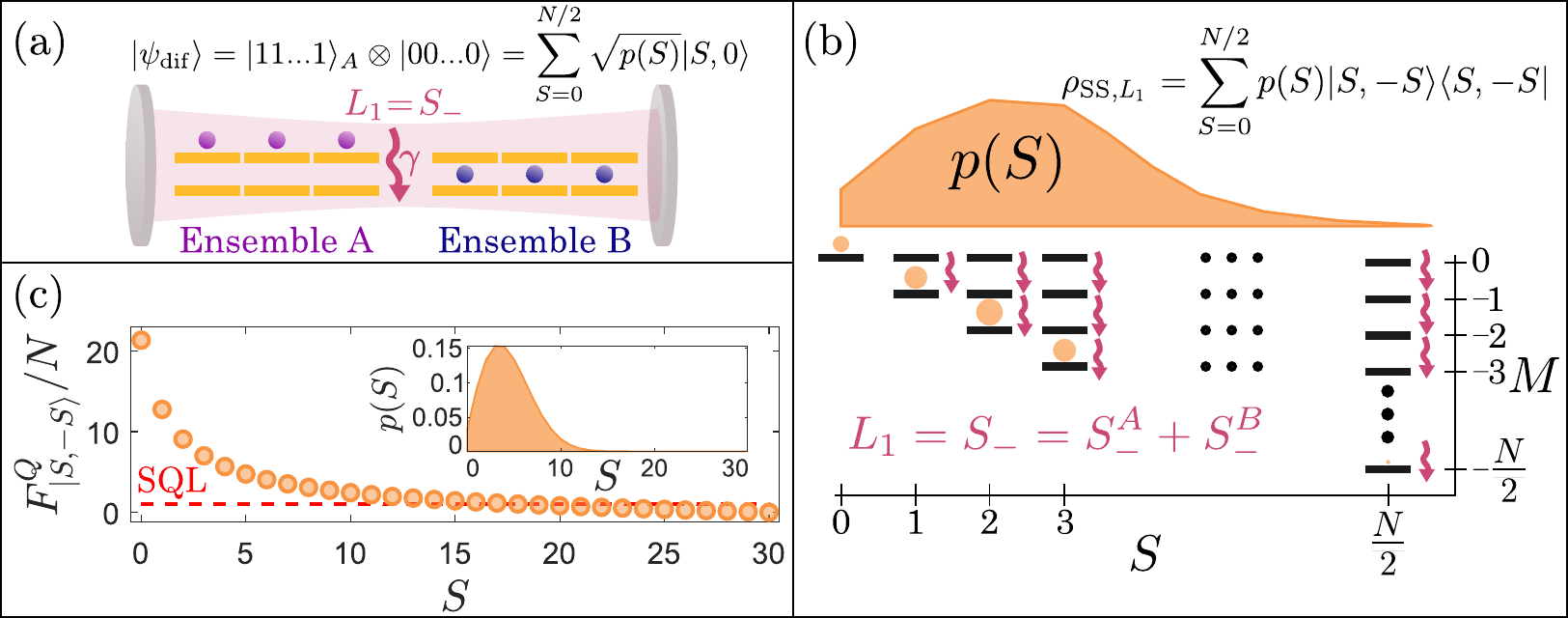}
\caption{\textbf{Two ensembles of qubits under collective decay:} \textbf{(a)} The system is initially prepared with all qubits in $A$ pointing up and all qubits in $B$ pointing down. This state can be written in the total angular momentum basis as $\ket{\psi_{\rm dif}}=\sum_{S=0}^{N/2}\sqrt{p(S)}\ket{S,0}$, where the distribution $p(S)$ is defined by the Clebsch-Gordan coefficients $p(S) = \vert \langle S,0 \vert N/4, +N/4, N/4, -N/4\rangle \vert^2$. We consider the two ensembles to be inside an optical cavity such that they can undergo collective decay characterized by the jump operator $L_1 = S_- = S_-^A + S_-^B$ and rate $\gamma$.
\textbf{(b)} The collective decay jump operator $L_1 = S_-$ causes the population on each sector of $S$ to decay into the state $\vert S,-S\rangle$ with the final population in each of these states given by $p(S)$ as indicated by the size of the circles. The steady state corresponds to a mixture of all such states $\rho_{{\rm SS},L_1} = \sum_{S=0}^{N/2}p(S) \vert S,-S\rangle \langle S \vert$, since the initial state does not have any coherences of the form $\ket{S,-S} \bra{S',-S'}$ for $S\neq S'$ (see main text). 
\textbf{(c)} The quantum Fisher information of each state $\ket{S,-S}$ is plotted as a function of $S$ for $N=60$ showing a rapid decay as $S$ increases. The inset shows the distribution $p(S)$ as a function of $S$, which peaks close to $\sqrt{N}$ as discussed in the main text. \crr{The red dashed line represents the SQL}. 
 }
\label{Fig3TwoEnsemble}
\end{figure*}

\section{Two ensembles of qubits used for differential sensing} \label{SecIV}  
In this section, we further benchmark the expressions in Eqs.~\eqref{eq:result1}-\eqref{eq:result2} for larger systems by generalizing the previous example to consider $N$ qubits. The qubits are separated into two ensembles, $A$ and $B$, containing $N_A$ and $N_B$ qubits, respectively. Where $N_A+N_B=N$. First, let's consider we initially prepare the two ensembles in a given state $\bm{\rho}_0$, and let them evolve through collective emission with the jump operator $L_1 = S_- =S^A_- + S^B_-$ and rate $\gamma$, where we define the collective spin ladder operators for each ensemble as $S_{\pm}^{A/B} =\frac{1}{2} \sum_{n \in A/B}\sigma_\pm^n$. It was recently shown that for two subensembles of equal size $N_A = N_B = N/2$, the steady state of this system can be used to estimate a phase $\phi$, accumulated under the unitary evolution described by the operator $U_{\rm G}= e^{-i \phi G}$, with generator $G = S_z^A - S_z^B$,
with a sensitivity surpassing the standard quantum limit~\cite{Raphael2025}. Here, we have defined $S_{\alpha}^{A/B} =\frac{1}{2} \sum_{n \in A/B}\sigma_\alpha^n$, for $\alpha=x,y,z$.  The initial state considered was $\vert \psi_{\rm dif} \rangle = \vert 11 \dots 1 \rangle_A \otimes \vert 00 \dots 0\rangle_B$, i.e., all atoms in $A$ pointing up and in $B$ down or vice versa. Schematics of this setup are presented in Fig.~\ref{Fig3TwoEnsemble}(a). In what follows, we find an analytical expression for the steady state generated through the evolution under $L_1$ for this initial state, using our formalism.

Similar to the two-qubit case, we can use Eq.~\eqref{eq:Albert} to predict the structure of the steady state. For the case $N_A=N_B$ and a Liouvillian defined by $L_1 = S_-$ and $H=0$, we find the kernel to have $n=(N/2+1)^2$ vectors $\bm{\mathcal{V}}_j$ of the form:

\begin{eqnarray}
&&\mathcal{V}_j = \ket{S,-S} \bra{S,-S},\quad 1 \leq j \leq N/2 +1\nonumber \\
&&\mathcal{V}_j = \ket{S,-S} \bra{S',-S'}, \quad S \neq S', \quad j > N/2+1,
\label{vecsUnb}
\end{eqnarray}
where $\ket{S,M}$ are the Dicke states defined by $S^2 \ket{S,M} = S(S+1) \ket{S,M}$ and $S_z \ket{S,M} = M \ket{S,M}$. Here, the operators are defined by $S_\alpha = S_\alpha^A+S_\alpha^B$ with $\alpha=x,y,z$, and $S^2 = S_x^2 + S_y^2 + S_z^2$. The eigenvalues are given by $S=0,1,\dots, N/2 +1$, and $M=-S,-S+1,\dots, S-1, S$. In Fig.~\ref{Fig3TwoEnsemble}(b), we provide a graphic depiction of these states. We note that the structure in Eq.~\eqref{vecsUnb} is the same as for the two-qubit case, the first $N/2+1$ vectors $\bm{\mathcal{V}}_j$ correspond to the dark states of the jump operator $L_1$, namely, $L_1 \ket{S,-S}=0$, while the remaining vectors correspond to the coherences between these dark states. For the first $N/2+1$ vectors, the corresponding vectors $\bm{\mathcal{U}}_j$ are given by $\mathcal{U}_j = \sum_{M=-S}^S \ket{S,M}\bra{S,M}$, where the index $j$ takes the value $j=S+1$, while for the remaining vectors ($j>N/2+1$) we require $\mathcal{U}_j = \sum_{M'=-S'}^{S'} \alpha_{S,S,M'}\ket{S,M'-(S-S')}\bra{S',M'}$, where we considered $S>S'$. $\alpha_{S, S', M'}$ are coefficients depending on all three quantum numbers.

We can explicitly compute the steady state by writing the initial state in the Dicke basis. We note that $\ket{\psi_{\rm dif}} = \ket{N/4,+N/4,N/4,-N/4}$ in the product basis $\ket{S_A,M_A,S_B,M_B}$. It follows that $\ket{\psi_{\rm dif}} = \sum_{S=0}^{N/2} C_S \ket{S,0}$, where $C_S=\bra{S,0}\ket{N/4,+N/4,N/4,-N/4}$ is the Clebsch-Gordan coefficient for a given $S$. We provide explicit expressions for these coefficients in \hyperref[app_E]{Appendix E}. Then, the initial density matrix can be written as $\rho_0 = \sum_{S,S'} C_S C_{S'}^* \ket{S,0}\bra{S',0}$. Using the expressions for $\bm{\mathcal{U}}_j$, we explicitly find the overlaps to be $\bm{\mathcal{U}}_j^\dagger \bm{\rho}_0={\rm tr}(\mathcal{U}_j^{\dagger} \rho_0) = \vert C_S \vert^2$ for $j\leq N/2+1$ and ${\rm tr}(\mathcal{U}_j^{\dagger} \rho_0) = 0$ for $j>N/2+1$. The latter overlaps disappear since the initial state only has coherences between states with the same $M=M'=0$, while $\bm{\mathcal{U}}_j$ for $j>N/1+1$ only contains coherences between states with $M\neq M'$, with the difference given by $M= M'-(S-S')$. Combining these results, and using Eq.~\eqref{eq:Albert}, the steady state of the system can be explicitly written as:

\begin{equation}
    \rho_{{\rm SS},L_1} = \sum_{S=0}^{N/2} p(S) \ket{S,-S} \bra{S,-S},\,
    \label{ssUnb}
\end{equation}
where we have defined $p(S) = \vert C_S \vert^2$. The locations of the distribution maxima scale as $\propto \sqrt{N}$~\cite{Raphael2025} [see Fig.\ref{Fig3TwoEnsemble}(c)]. For the initial state studied here, the jump operator $L_1$ causes the population in each sector of fixed $S$ to decay from state $\ket{S,0}$ to state $\ket{S,-S}$ along the Dicke ladder. This is illustrated in Fig.~\ref{Fig3TwoEnsemble}(b). We note that the structure of the steady state follows the attractor structure discussed in Sec.~\hyperref[SecII]{II}. In each trajectory (an individual run of an experiment, for example), the system will evolve into one of the attractor states $\ket{S,-S}$ with probability $p(S)$. We represent the averaging over many of these trajectories with the mixed-state density matrix $\rho_{{\rm SS},L_1}$.

\begin{figure*}[th!]
\includegraphics[width=0.98\textwidth]{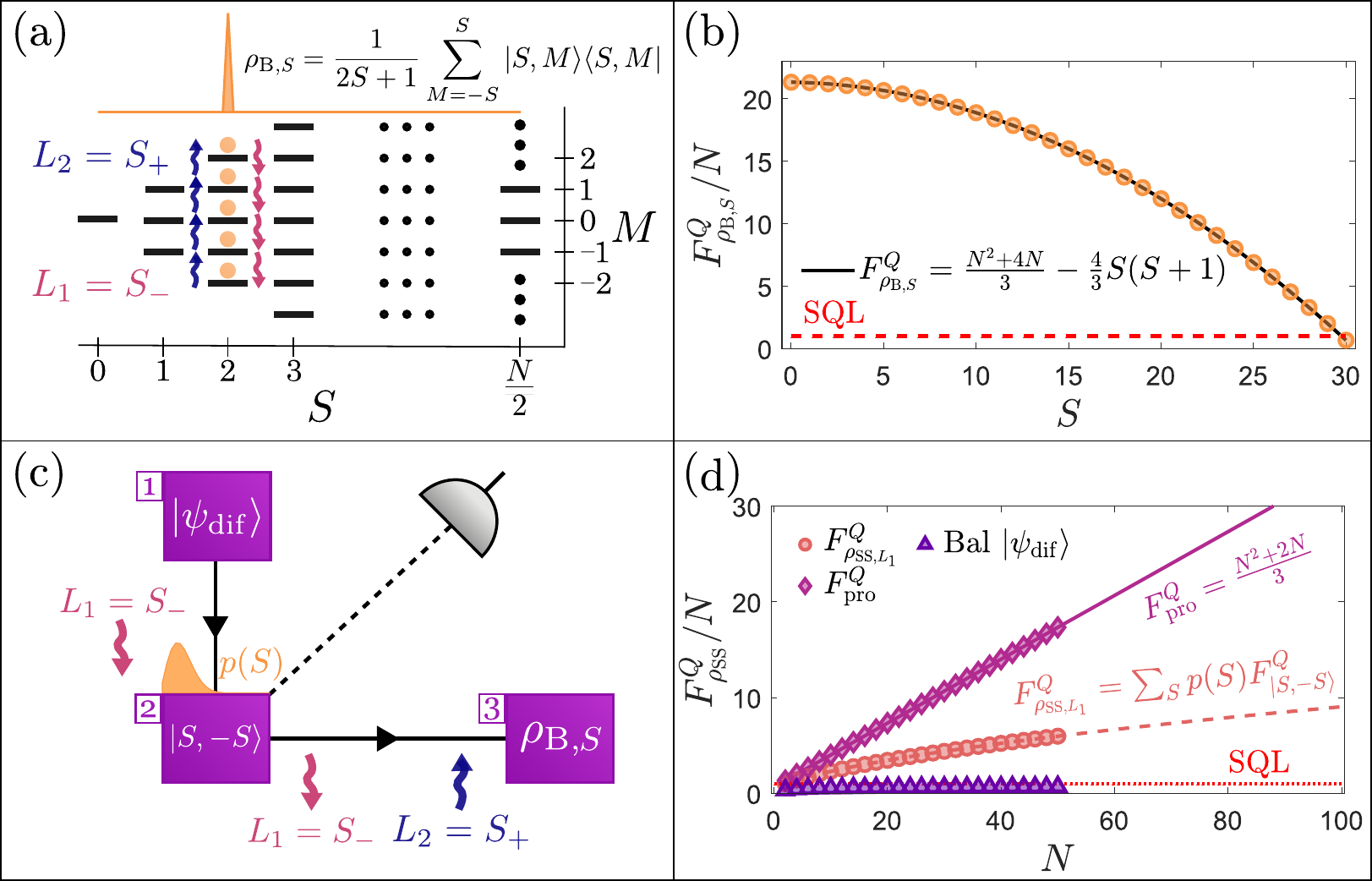}
\caption{\textbf{Enhancing the QFI using balanced decay processes:} \textbf{(a)} The action of balanced collective jump operators $L_1 = S_-$ and $L_2 = S_+$ causes an initial Dicke state $\ket{S,M}$ to evolve into state $\rho_{{\rm B},S}$ which is an equal mixture of all Dicke states $\ket{S,M}$ with fixed $S$ but different $M$.
\textbf{(b)} The quantum Fisher information of each state $\rho_{{\rm B},S}$ is plotted as a function of $S$ for $N=60$. The numerical values (markers) computed with Eq.~\eqref{eq:QFI2} agree with the analytical expression found in \hyperref[app_E]{Appendix E}. \crr{The red dashed line represents the SQL.} 
\textbf{(c)} Schematics of a single run of the protocol described in the main text. First, the system is initialized in the state $\ket{\psi_{\rm dif}}$. Subsequently, it decays collectively with the jump operator $L_1$ until it reaches the steady state $\ket{S,-S}$ with probability $p(S)$. Finally, we turn on the $L_2$ process and let the system evolve under the action of the balanced jump operators $L_1$ and $L_2$, reaching the steady state $\rho_{{\rm B},S}$. Additionally, if we wanted to do some post-processing to only keep evolutions that lead to low values of $S$ (high QFI), we could include measurements during the first evolution (dotted line). In a cavity setup, this could be done by continuously collecting the photons leaking out of the cavity while the system evolves under $L_1$. By doing so, we could know the specific value of $S$ and then discard the experimental runs corresponding to a large $S$ value.
\textbf{(d)} Quantum Fisher information of different steady states as a function of the system size $N$. The markers represent steady states obtained numerically with Eq.~\eqref{eq:result2} using $\epsilon = 0.00001$. The solid line is the analytic expression for the QFI obtained for the protocol after many runs, while the dashed curve represents the QFI for the state $\rho_{{\rm SS},L_1}$ obtained for evolution under $L_1$ only [Eq.~\eqref{ssUnb}]. Details on how these two curves are obtained are provided in \hyperref[app_E]{Appendix E}. Purple triangles correspond to the steady state obtained when the initial state $\ket{\psi_{\rm dif}}$ is evolved under both $L_1$ and $L_2$. \crr{The red dotted line represents the SQL.}}
\label{Fig4TwoEnsemble}
\end{figure*}

 Now that we have found an expression for the steady state, we can characterize its entanglement properties. \crr{As we discuss below, the quantum states most suitable for metrological tasks are characterized by a large entanglement depth and usually possess fairly limited bipartite entanglement, making a characterization in terms of entanglement depth more suitable in the metrological context.} The quantum Fisher Information (QFI) can be used as a metric to quantify how sensitive a quantum state is to changes of a parameter $\phi$~\cite{Pezze2018,Huang2024}. The higher the QFI, the higher the sensitivity of that state. For a system of $N$ qubits prepared in a separable state (unentangled), when considering a generator of the form $G = S_z^A - S_z^B$, the maximum value the QFI can take is $N$ \footnote{This result can be generalized to any generator of the form $G = \sum_{j=1}^N \bm{n}\cdot \sigma_{\bm {n}}^j/2$, where $\bm{n}$ is a normalized vector and $\sigma_{\bm {n}}$ is the Pauli vector in the $\bm{n}$ direction \cite{Pezze2009entanglement}}. This bound is known as the standard quantum limit (SQL). To surpass this limit, we require the state to have some degree of entanglement. For an entangled state, the maximum QFI is given by $N^2$. \crr{A state saturating this bound, referred to as the Heisenberg limit, necessarily contains genuine $N$-particle entanglement~\cite{Pezze2009entanglement}.} Additionally, a state is said to exhibit Heisenberg scaling when its QFI scales $\propto N^2$. As mentioned before, we are interested in signals $\phi$ that couple differentially to the population in the two ensembles, i.e., the evolution of the state concerning the field producing the signal is modeled by the generator $G = S_z^A - S_z^B$. In what follows, we denote the QFI corresponding to that encoding by $F^Q_{\rho}$, for any arbitrary state $\rho$, and by $F^{Q}_{\ket{\psi}}$ for pure states $\ket{\psi}$. The details on the mathematical expressions we use to compute the QFI are provided in the \hyperref[app_D]{Appendix D}. 

\crr{The QFI allows us to characterize the entanglement properties of a state even further. Let's assume we divide our $N$ qubits into $r$ groups of $k$ qubits each. (we assume that $N$ is divisible by $k$ \footnote{If $N$ is not divisible by $k$, the bound in Eq.~\eqref{depth1} can be generalized to $F^Q_{\rho} \leq \lfloor N/k\rfloor k^2 + N-\lfloor N/k\rfloor k$, where $\lfloor N/k \rfloor$, rounds down to the greatest integer that is less than or equal to $N/k$.}). The qubits within each group are allowed to be entangled with each other, but not with those in other groups. The density matrix of such a state is given by $\rho = \rho_1 \otimes \rho_2 \otimes \dots \otimes \rho_r$. We consider each group of $k$ qubits to be in a maximally $k$-particle entangled state with QFI given by $F_{\rho_{j}}^Q = k^2$. Using the properties of the QFI for separable states, we have that $F_\rho^Q=\sum_{j=1}^r F_{\rho_j}^Q = r k^2 = Nk$. Then, we can characterize the entanglement depth of a state with the bound~\cite{multiLaskowski,multiToth}

\begin{equation}
F^Q_{\rho} \leq Nk\,.
    \label{depth1}
\end{equation}

Note that for $k=1$ we recover the expression for the SQL. The way to utilize this bound is as follows: any state that breaks this bound necessarily contains at least an entanglement depth of $k+1$~\cite{SorensenMolmerDepth}, i.e., at least $k+1$ particles are genuinely entangled. The converse is not true, namely, a state with entanglement depth of $l$ with $l>k$ might still have a QFI within the bound above. Since the QFI quantifies the potential metrological utility a state has, states with large metrological utility necessarily have large entanglement depth. In what follows, we characterize the entanglement depth of the steady state in Eq.~\eqref{ssUnb}.}

 In Fig.~\ref{Fig3TwoEnsemble}(c), we show the QFI for each Dicke state $\ket{S,-S}$ for $N=60$. The QFI decays rapidly as $S$ increases; hence, to maximize the steady state entanglement \crr{depth}, one would like to have an initial state that has the largest possible overlap with the $\ket{0,0}$ state. As shown in the inset, the distribution $p(S)$ for the initial state $\ket{\psi_{\rm dif}}$ is skewed towards smaller values of $S$. \crr{Since states with low $S$ have large entanglement depth,} the steady state in Eq.~\eqref{ssUnb} has a large QFI even when it corresponds to a mixed state. \crr{This highlights the important role of the initial state choice in determining the entanglement properties of the resulting steady state. For example, if we had chosen the initial state to be the state $\ket{N/2,N/2}$ instead of $\vert \psi_{\rm dif}\rangle$, the steady state reached through the action of the same Liouvillian would be the separable (not entangled) state $\vert N/2, -N/2 \rangle$. In that case, not only is the steady state unentangled, but no entanglement is developed at all in the dynamics towards the steady state, as shown in Refs.~\cite{SusanneSeparability,Porras2013,SussaneDriven2014,Rosario2025,bassler2025absence}.}
 
 In Fig.~\ref{Fig4TwoEnsemble}(d) we show the QFI of $\rho_{{\rm SS},L_1}$ as a function of $N$, where we note that the numerical results obtained with Eq.~\eqref{eq:result2} (circles) agree with the simplified expression for the QFI of these states that we find in \hyperref[app_E]{Appendix E} (dashed line). Although $\rho_{\mathrm{SS},L_1}$ exhibits a large QFI (\crr{significantly above the SQL}), it does not achieve Heisenberg scaling; namely, the QFI does not scale proportionally to $N^{2}$ for large $N$. This raises the question of whether the previously described protocol can be incorporated into a broader scheme to generate a different steady state with improved QFI scaling. In what follows, we propose employing the balanced decay processes studied in the two-qubit case to enhance the QFI of the resulting steady state.

Again, we consider the Liouvillian generated by the two jump operators $L_1 = S_- = S_-^A+S_-^B$ and $L_2=S_+=S_+^A+S_+^B$ with balanced rates $\gamma_1=\gamma_2=\gamma$. This setup allows us to use Eq.~\eqref{result:specialcase}. For the case $N_A=N_B=N/2$, we find that the system has $n=N/2 +1$ eigenvectors $\bm{\rho}_{{\rm SS},j}$, which in matrix form are given by:

\begin{eqnarray}
&&\rho_{{\rm SS},j} = (-1)^{j-1} \sqrt{2j -1}\, \rho_{{\rm B},j-1}\,, \nonumber \\
&&\rho_{{\rm B},S} = \sum_{M=-S}^S \frac{1}{2S+1}\ket{S,M} \bra{S,M}\,.
\label{balancedDicke}
\end{eqnarray}

As in the two-qubit case, each density matrix $\rho_{{\rm B},S}$ corresponds to an equal mixture of all states with a fixed eigenvalue $S$, reflecting the conservation of total spin length implied by $[L_1,S^2] = [L_2,S^2] = 0$. In Fig.~\ref{Fig4TwoEnsemble}(b), we show the QFI of the density matrices $\rho_{{\rm B},S}$, which show a considerably slower decay of the QFI as a function of $S$ compared to the case of the Dicke states $\ket{S,-S}$ [Fig.~\ref{Fig3TwoEnsemble}(c)]. We find that the QFI of these states follows $F^Q_{\rho_{{\rm B},S}}= \frac{N^2 +4N}{3}-\frac{4}{3}S(S+1)$ (see \hyperref[app_E]{Appendix E} for the derivation). It is important to note that due to the factor of $\sqrt{2S+1}$ in $\bm{\rho}_{{\rm SS},j}$, for an arbitrary initial state $\bm{\rho}_0$, the weight $c_j = \bm{\rho}_{{\rm SS},j}^\dagger \bm{\rho}_0$ for large $S$ (low QFI) can be large even if the overlap $\bm{\rho}_{{\rm B},S}^\dagger \bm{\rho}_0$ is small, leading to an overall smaller QFI. For example, if we start our dynamics in the state $\ket{\psi_{\rm dif}}$, the decay under the balanced operators $L_1$ and $L_2$ yields a steady state with lower QFI than the one generated exclusively with decay under $L_1$ as shown in panel (d) of Fig.~\ref{Fig4TwoEnsemble} (purple triangles). 

\begin{figure*}[th!]
\includegraphics[width=0.98\textwidth]{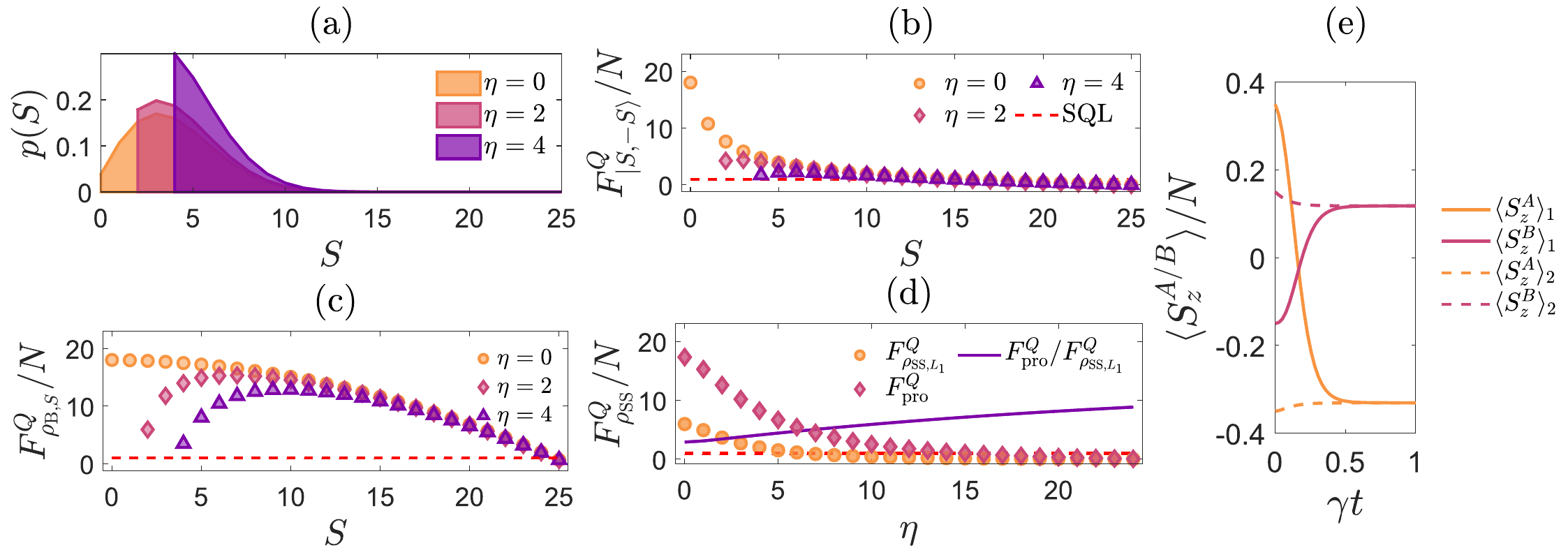}
\caption{\textbf{Effects of population imbalance:} \textbf{(a)} The distribution $p(S)$ corresponding to the initial state $\ket{\psi_{\rm dif}}$ is shown for different imbalance values $\eta$.
\textbf{(b)} The quantum Fisher information of each state $\ket{S,-S}$ is plotted as a function of $S$. Different markers correspond to different values of the imbalance. \crr{In this and the next two panels, the red dashed line represents the SQL.}  
\textbf{(c)} The quantum Fisher information of each state $\rho_{{\rm B},S}$ is plotted as a function of $S$. Different markers correspond to different values of the imbalance.  
\textbf{(d)} Quantum Fisher information of different steady states as a function of the imbalance $\eta$. All the data was obtained with Eq.~\eqref{eq:result2} using $\epsilon = 0.00001$. The circle markers signal the QFI of the state $\rho_{{\rm SS},L_1}$ obtained after the evolution under $L_1=S_-$, while the diamonds represent the QFI of the state at the end of the protocol. The solid purple line corresponds to the ratio of these two quantities.
\textbf{(e)} Dynamics of the observables $\langle S_z^A \rangle$ and $\langle S_z^B \rangle$ when the system evolves under the jump operator $L_1$ and rate $\gamma$ into the steady state $\rho_{{\rm SS},L_1}$. Orange curves correspond to $\langle S_z^A \rangle$, while the pink ones correspond to $\langle S_z^B \rangle$. The solid lines represent the dynamics when the system is initialized in $\ket{\psi_{\rm dif}}$, while the dashed lines correspond to the initial state $\ket{\psi_{{\rm dif}_2}}$. For all panels, we consider $N=50$.
 }
\label{Fig5}
\end{figure*}

Although balanced processes are not guaranteed to generate useful steady states for arbitrary initial states, if we manage to initially prepare the system in any state with a well defined eigenvalue $S$, the steady state would be given by $\rho_{{\rm B},S}$ [Eqs.\eqref{balancedDicke} and \eqref{result:specialcase}] which displays a large QFI, especially for small values of $S$. For example, schematics on how a Dicke state $\ket{S,M}$ evolves into $\rho_{{\rm B},S}$ under the simultaneous action of $L_1$ and $L_2$ are shown in Fig.~\ref{Fig4TwoEnsemble}(a). Motivated by this, we propose the following protocol to generate optimal initial states for the evolution under $L_1$ and $L_2$, which we describe schematically in Fig.~\ref{Fig4TwoEnsemble}(c). First, we prepare the system in $\ket{\psi_{\rm dif}}$ and let the system decay under collective decay with jump operator $L_1$ with rate $\gamma_1$ until it reaches the steady state. As described by Eq.~\eqref{ssUnb}, in each run of the experiment, the system will evolve into a Dicke state $\ket{S,-S}$ with probability $p(S)$ \crr{(see \hyperref[app_C]{Appendix C})}. Finally, we turn on the second collective jump operator $L_2$ with a balanced rate $\gamma_2=\gamma_1=\gamma$, and evolve under both jump operators until we reach the steady state $\rho_{{\rm B},S}$. Additionally, one could include measurements after the first step to post-select only the runs when the dynamics lead to a state $\ket{S,-S}$ with a low value of $S$ (high QFI). In a cavity QED setup, this could be done by continuously collecting all photons that leak out of the cavity during the entire evolution under $L_1$, or in a different setup, through a collective projective measurement of $S_z = S_z^A+S_z^B$.

If we don't consider any post-selection, we can quantify the QFI for the protocol by the expression $F^Q_{\rm pro} = \sum_{S=0}^{N/2} p(S) F^Q_{\rho_{{\rm B},S}}$. We find that the QFI is explicitly given by $F^Q_{\rm pro} = \frac{N^2+2N}{3}$ (see \hyperref[app_E]{Appendix E} for details), which shows Heisenberg scaling and a large enhancement with respect to the QFI of state $\rho_{{\rm SS},L_1}$, as we show in Fig.~\ref{Fig4TwoEnsemble}(d), indicating the potential utility of these states for differential sensing. We note that the numerical data generated with Eq.~\eqref{eq:result2} (pink diamonds) agrees with the analytically predicted values for $F^Q_{\rm pro}$, showing again the utility of Eq.~\eqref{eq:result2}.

Finally, we explore how an imbalance of the populations in the two ensembles can affect the entanglement properties of the steady states. We set $N_A = N/2 + \eta$ and $N_B = N/2-\eta$, with $\eta$ being a positive integer, and $N_A + N_B=N$. We consider again the initial state $\vert \psi_{\rm dif} \rangle = \vert 11 \dots 1 \rangle_A \otimes \vert 00 \dots 0\rangle_B$, but now since $N_A \geq N_B$, the expansion in the Dicke states is given by $\ket{\psi_{\rm dif}} = \sum_{S=\eta}^{N/2} C_S^\eta \ket{S,\eta}$, where $C_S^\eta=\bra{S,\eta}\ket{N/4+\eta/2,+N/4 +\eta/2,N/4-\eta/2,-N/4 + \eta/2}$. Again, we define $p(S) = \vert C_S^\eta \vert^2$. In Fig.~\ref{Fig5}(a), we show $p(S)$ for $N=50$ and $\eta=0,2,4$. Now, if we let the system evolve under $L_1 = S_-$, we find the steady state to have the same form as for $\eta=0$ [Eq.~\eqref{ssUnb}], with the difference that now the allowed values of $S$ are $\eta \leq S \leq N/2$, namely, $\rho_{{\rm SS},L_1} = \sum_{S=\eta}^{N/2} p(S) \ket{S,-S} \bra{S,-S}$. We should note, however, that since $N_A \neq N_B$, the structure of $G = S_z^A- S_z^B$ changes, and, consequently, the QFI of the states $\ket{S,-S}$ depends on the value of $\eta$. In Fig.~\ref{Fig5}(b), we show how the QFI decreases rapidly as a function of $\eta$ for these Dicke states. This decreasing behavior of the QFI, added to the fact that $p(S)$ is skewed towards larger values of $S$ when $\eta$ increases, causes $\rho_{{\rm SS},L_1}$ to decrease its QFI significantly as a function of $\eta$. This behavior is shown in Fig.~\ref{Fig5}(d).

An interesting consequence of the form of $\rho_{{\rm SS},L_1}$ is that the initial state $\ket{\psi_{{\rm dif}_2}} = \vert 00 \dots 0 \rangle_A \otimes \vert 11 \dots 1\rangle_B$, where all qubits in $A$ are pointing down and all in $B$ are pointing up, yields the same steady state $\rho_{{\rm SS},L_1}$ as the initial state $\ket{\psi_{\rm dif}}$. This is caused by the fact that $p(S) = (C_S^\eta)^2$ and $\vert \bra{S,\eta}\ket{N/4+\eta/2,+N/4 +\eta/2,N/4-\eta/2,-N/4 + \eta/2}\vert^2 = \vert \bra{S,-\eta}\ket{N/4+\eta/2,-N/4 -\eta/2,N/4-\eta/2,N/4 - \eta/2}\vert^2$ due to the symmetry properties of the Clebsch-Gordan coefficients. In Fig.~\ref{Fig5}(e), we consider the evolution of a system with $N=50$ and $\eta =10$ under the action of the jump operator $L_1=S_-$ only. We integrate the dynamics for the two initial states $\ket{\psi_{\rm dif}}$ and $\ket{\psi_{{\rm dif}_2}}$ and confirm that they reach the same steady state $\rho_{{\rm SS},L_1}$. In order to reach the same steady state, these two initial conditions undergo very different dynamics. When the largest ensemble starts pointing down, the system reaches the steady state very fast, as the initial state is already close to the steady state. On the other hand, when the largest ensemble starts pointing up, both ensembles invert their populations almost entirely during the transient dynamics. Similar dynamics of collective decay when a fraction of the spins are initially excited have been studied before in the context of waveguides~\cite{Fasser:24}. This illustrates that the expressions derived here, accurately determine the steady state of the system for a given initial state, regardless of the complexity of the transient dynamics.

Now, we consider the protocol described in Fig.~\ref{Fig4TwoEnsemble}(c) for the imbalanced case. In each run of the experiment, the steady state will be the state $\rho_{{\rm B},S}$ with probability $p(S)$, with the caveat that now $\eta \leq S \leq N/2$. Just as for the Dicke states, the QFI of density matrices $\rho_{{\rm B},S}$ decreases as $\eta$ increases, as shown in Fig.~\ref{Fig5}(c). However, we note that the effect of $\eta$ is less abrupt in these states, and consequently, we expect $F^Q_{\rm pro}$ to be considerably larger than $F^Q_{\rho_{{\rm SS},L_1}}$. This is confirmed in Fig.~\ref{Fig5}(d), where not only we find that $F^Q_{\rm pro}/F^Q_{\rho_{{\rm SS},L_1}} > 1$ for all $\eta$, but also that the ratio between them grows as a function of $\eta$, meaning that the protocol described here can also help mitigate the loss of QFI due to the imbalance of population between the ensembles.

As a final note, in Figs.~\ref{Fig3TwoEnsemble}, ~\ref{Fig4TwoEnsemble}, and ~\ref{Fig5}, we used Eq.~\eqref{eq:result2} to obtain the steady states of each process numerically. For many of the analyses, we studied the dependence on system size $N$ or imbalance $\eta$, which means that for each data point the Hilbert space size changes and quantities such as $T$ [Eq.~\eqref{eq:result1}] or $\bm{\mathcal{V}}_j$ and $\bm{\mathcal{U}}_j$ [Eq.~\eqref{eq:Albert}] have to be recalculated each time. As discussed at the end of Sec.~\hyperref[SecII]{II}, for these cases, if we can efficiently compute $\left(\mathbb{I} - \frac{1}{\epsilon}\mathcal{L}\right)^{-1}\bm{\rho}_0$ without explicitly inverting $\left(\mathbb{I} - \frac{1}{\epsilon}\mathcal{L}\right)^{-1}$, Eq.~\eqref{eq:result2} can become significantly more efficient than the other expressions, as no prior computation and storage of quantities like $T$ are needed.

\section{Conclusions} \label{SecV} 
We have presented a set of expressions, Eqs.~\eqref{eq:result1}-\eqref{eq:result2}, that enable the determination of the steady state of open quantum systems governed by the Lindblad equation without the need for explicit integration of the dynamics. As demonstrated here, Liouvillians with collective jump operators impose symmetries that allow the system to support multiple steady states. Although we have restricted our examples to purely dissipative evolution ($H = 0$), the expressions introduced here can also be used to determine the steady states of systems with $H \neq 0$, as shown in the Supplemental Material~\cite{supplemental}.

In the examples considered, tailoring the overlap of the initial state with elements of the Liouvillian’s kernel can enhance desirable features, such as quantum entanglement, in the resulting steady state. This connection become particularly clear because analytical expressions for the structure of the Liouvillian's null space are accessible and exhibit a close relation to the system’s symmetries. The role of the initial state is particularly transparent for cases where $\mathcal{L} = \mathcal{L}^{\dagger}$ [see Eq.~\eqref{result:specialcase}], since the elements of the kernel can be spanned by valid density matrices onto which the initial state is projected [Fig.~\ref{Fig1Sch}(b)]. We exploit these capabilities to propose a protocol for generating metrologically useful states by evolving the system under two balanced collective decay channels. As noted earlier, such balanced dissipation can be realized in cavity QED setups with two dressing beams~\cite{luo2024realization,luo2025hamiltonian}, enabling the experimental exploration of the steady-state properties of this type of Liouvillians.

For more complex systems, analytical expressions may not be available, and isolating a single overlap to maximize a desired property may not be possible. Nonetheless, all of the presented expressions, particularly Eq.~\eqref{eq:result2}, can be employed for numerical optimization over a set of initial conditions without prior knowledge of the system's symmetries. We note that these expressions are valid only when the steady-state limit $\lim_{t \rightarrow \infty} \rho(t)$ exists; hence, they cannot be applied to purely unitary (Hamiltonian-only) dynamics. In Fig.~\ref{Fig5}(e), we confirmed that our approach accurately predicts the steady state regardless of the complexity of the transient dynamics. However, if the primary interest lies in transient dynamics, the present formalism is not functional. On the other hand, as shown in the Supplemental Material~\cite{supplemental}, when the goal is solely to determine the long-time behavior, evaluating the expressions presented here is often more numerically efficient than simulating the full time evolution for long times.

\begin{acknowledgments}
We thank Amit Vikram and Eliot Bohr for useful feedback on the manuscript.  This material is based upon work supported by the Vannevar Bush Faculty Fellowship,  the NSF JILA-PFC PHY-2317149 and OMA-2016244 (QLCI), the U.S. Department of Energy, Office of Science, National Quantum Information Science Research Centers, Quantum Systems Accelerator, and NIST. 

\end{acknowledgments}
 
\bibliography{refs}{}

\appendix

\section{Derivation of the steady state formula presented in main text }\label{app_A}

The density matrix $\rho$ can be represented by a $\mathcal{D} \times \mathcal{D}$ matrix.  Consequently, $\mathcal{L}$ can  be also represented by a $\mathcal{D}^2 \times \mathcal{D}^2$ matrix, and $\bm{\rho}$ by a $\mathcal{D}^2 \times 1$ vector. 

To derive a new expression for the steady-state solution of Eq.\eqref{eq:1}, we first introduce  the Laplace transform (LT) which is used to solve systems of linear ordinary differential equations (ODEs). For any function $f(t)$ we define its LT [denoted $F(s)$] as~\cite{ARFKEN2013963}:

\begin{equation}
    \hat{L}\{f(t)\}(s)=F(s) = \int_0^{\infty}e^{-st}f(t)dt\,.
    \label{eqEM:1}
\end{equation}

This integral transform is deeply connected to Green's functions~\cite{MITRA01011990}, which are widely used in almost any physics field.  It has also been implemented in the treatment of time-fractional physics problems \cite{ALQURAN2020103667}, and has been incorporated in the treatment of open quantum systems evolving under Eq.~\eqref{eq:1} in the context of thermodynamic currents~\cite{manzano2014,Juzar2021}. A key consequence of Eq.~\eqref{eqEM:1} is:

\begin{equation}
\hat{L}\left\{ \frac{df(t)}{dt} \right\}(s)= s F(s) - f(0)\,,
\label{eqEM:2}
\end{equation}
where the initial value $f(0)$ is incorporated explicitly. Now, we define $\bm{R}(s) = \hat{L}\{\bm{\rho}(t)\}(s)$ as the LT of the vectorized density matrix $\bm{\rho}(t)$. Applying the Laplace transform to both sides of Eq.~\eqref{eq:1} and doing some algebra, we obtain

\begin{equation}
(s\mathbb{I} - \mathcal{L}) \bm{R}(s) = \bm{\rho}_0\,,
\label{eqEM:3}
\end{equation}
where $\mathbb{I}$ denotes the $\mathcal{D}^2\times \mathcal{D}^2$ identity and $\bm{\rho}_0$ is the vectorized density matrix at time $t=0$. Here, we have considered a time-independent Liouvillian and used the linear property of the LT. We can express the matrix  $\mathcal{L}$ using its singular value decomposition (SVD) as $\mathcal{L} = U \Sigma V^{\dagger}$, where $\Sigma$ is a diagonal matrix and both $U$ and $V$ are unitary matrices. In terms of these matrices, we can rewrite Eq.~\eqref{eqEM:3} as:

\begin{equation}
\bm{R}(s) = V\left(sU^{\dagger}V - \Sigma\right)^{-1} U^{\dagger} \bm{\rho}_0
\label{eqSVD1} \,,
\end{equation} where we have assumed that $sU^{\dagger}V - \Sigma$ is invertible even if $\Sigma$ has singular values equal to zero. Since we are interested in the steady state solutions, namely, the limit $t \rightarrow \infty$, we can use the final value theorem of the LT, which states that~\cite{schiff2013laplace}: 

\begin{equation}
    \lim_{t\rightarrow \infty} f(t) = \lim_{s\rightarrow0} s F(s)\,.
    \label{eqEM:5}
\end{equation}
In Eq.~\eqref{eqEM:5}, we assumed that both limits exist, which means that, as discussed in Sec.~\hyperref[SecV]{V}, the expressions derived below are only valid when the system reaches a steady state for long times. We can use this theorem to relate $\bm{\rho}(t)$ to $s\bm{R}(s)$ for $t\rightarrow \infty$. Using Eqs.~\eqref{eqSVD1} and~\eqref{eqEM:5} we obtain:

\begin{equation}
\bm{\rho}_{\rm SS} = \lim_{s \rightarrow 0} s \bm{R}(s) = V \left(\lim_{s \rightarrow 0}s\left(s U^{\dagger}V - \Sigma \right)^{-1}\right) U^{\dagger}\bm{\rho}_0\,.
\label{eqSVD2}
\end{equation}

This equation can be simplified (see Supplemental Material~\cite{supplemental} for details) to the form:

\begin{equation}
\bm{\rho}_{\rm SS} = \sum_{j=1}^n \bm{\mathcal{U}}^{\dagger}_j  \bm{\rho}_0  \bm{\mathcal{V}}_j\,,
\label{eqSVD3}
\end{equation} where $n$ is the number of zero singular values in $\Sigma$. $\bm{\mathcal{U}}_j$ are the columns of $U$ corresponding to the singular value equal to zero. Similarly, $\bm{\mathcal{V}}_j$ are the columns of $V$ corresponding to a singular value equal to zero. To obtain the expression above, we have assumed that the sets $\bm{\mathcal{U}}_j$ and $\bm{\mathcal{V}}_j$ are biorthogonal, that is, $\bm{\mathcal{U}}_j^{\dagger} \bm{\mathcal{V}}_k = \delta_{jk} $. In \hyperref[app_B]{Appendix B}, we show step by step how such biorthogonal sets can always be constructed. By the definition of the SVD, the set $\bm{\mathcal{V}}_j$ corresponds to the null space of $\mathcal{L}$, and the set $\bm{\mathcal{U}}_j$ corresponds to the null space of $\mathcal{L}^{\dagger}$.  Eq.~\eqref{eqSVD3} is the result first reported in~\cite{Albert2014,Albert2016PRX} and also Eq.~\eqref{eq:Albert} in the main text.

In the more specific case where $\mathcal{L}$ is diagonalizable through a transformation $T^{-1} \mathcal{L} T = \Lambda$, where $\Lambda = \text{diag}(\lambda_1, \lambda_2, ...,\lambda_{\mathcal{D}^2})$ is a diagonal matrix that has, at least, one zero eigenvalue ($\lambda_j=0$), Eq.~\eqref{eqSVD2} becomes:

\begin{equation}
\bm{\rho}_{\rm{SS}} = \lim_{s \rightarrow 0}s \bm{R}(s) = T\left(\lim_{s\rightarrow 0} s (s\mathbb{I} - \Lambda)^{-1}\right)T^{-1}\bm{\rho}_0\,.
\label{eqEM:6}
\end{equation}

Since $\Lambda$ is diagonal, it follows that:
\begin{equation}
 s (s\mathbb{I} - \Lambda)^{-1} = \text{diag}\left(\frac{s}{s-\lambda_1},\frac{s}{s-\lambda_2},...,\frac{s}{s-\lambda_{\mathcal{D}^2}}\right)\,.
 \label{eqEM:7}
\end{equation} 
If we take the limit of this expression as $s\rightarrow 0$, we notice that the multiplicative factor of $s$ in Eq.~\eqref{eqEM:5} removes the divergence that $\lambda_j=0$ would cause. Without this term, the matrix $s(s\mathcal{I}-\Lambda)^{-1}$ would have diverging eigenvalues when the limit $s\rightarrow0$ is considered. In the case where $n$ of the $\mathcal{D}^2$ eigenvalues of $\mathcal{L}$ are zero, $\lim_{s\rightarrow 0} s (s\mathbb{I} - \Lambda)^{-1}$ is a diagonal matrix with $n$ eigenvalues exactly equal to 1, and with the remaining $\mathcal{D}^2-n$ eigenvalues being equal to zero. We can recast Eq.~\eqref{eqEM:6} in the form:

\begin{equation}
    \bm{\tilde{\rho}}_{\rm{SS}}  = \begin{pmatrix}\mathbb{I}_n & 0 \\ 0 &0 \end{pmatrix} \bm{\tilde{\rho}}_0 = \sum_{j=1}^n \left(\bm{e}_j^{\dagger} \bm{\tilde{\rho}}_0\right) \bm{e}_j\,,
    \label{eq:rotated}
\end{equation} where we have considered that $\lambda_1,\lambda_2,...\lambda_n=0$ and all other $\lambda_j$ ($j>n$) are non-zero. $\bm{\tilde{\rho}}_{\rm{SS}} = T^{-1}\bm{\rho}_{\rm SS}$ and  $\bm{\tilde{\rho}}_0=T^{-1} \bm{\rho}_0$ are the steady state and initial vectorized density matrices in the basis where $\mathcal{L}$ is diagonal, $\mathbb{I}_n$ is the $n$ by $n$ density matrix, and $\bm{e}_j$ are a set of $n$ orthonormal column vectors of the form $\bm{e}_j = (0,0,\dots0,1,0,\dots,0,0)^T$ where the only non-zero value is located in the $j$-th row. Note that each vector $\bm{e}_j$ has $\mathcal{D}^2$ rows.

Applying the transformation $T$ to both sides of Eq.~\eqref{eq:rotated} we find:

\begin{equation}
    \bm{\rho}_{\rm{SS}}  = \sum_{j=1}^n c_j \bm{\rho}_{{\rm SS},j},\, c_j = (\bm{\rho}_{{\rm SS},j})^{\dagger} (T^{-1})^{\dagger} T^{-1} \bm{\rho}_0\,,
    \label{eq:resultEndMat}
\end{equation}which corresponds to Eq.~\eqref{eq:result1}. Finally, without recasting $\mathcal{L}$ using an SVD or a similarity transformation, we can find an alternative expression for the steady-state density matrix by combining Eqs.~\eqref{eqEM:3} and ~\eqref{eqEM:5}:

\begin{equation}
\bm{\rho}_{\rm SS} = \lim_{s \rightarrow 0}s \bm{R}(s) = \left(\lim_{s\rightarrow 0} s (s\mathbb{I} - \mathcal{L})^{-1}\right)\bm{\rho}_0 \,.
\label{eqEM:8}
\end{equation}




This expression  is Eq.~\eqref{eq:result2} in the main text. We emphasize that the results in Eqs.~\eqref{eqSVD3},~\eqref{eq:resultEndMat}, and~\eqref{eqEM:8} are derived using the final value theorem, consequently, these expressions cannot be used to study oscillatory systems where the limit $\lim_{t\rightarrow \infty} \bm{\rho}(t)$ is not well defined.

\section{Step-by-step implementation of the steady-state calculation}\label{app_B}

Here we outline a numerical implementation of Eq.~\eqref{eq:Albert} for a given Hamiltonian H and set of jump operators $L_j$. First, we consider the Liouvillian superoperator 

\begin{eqnarray}
    \mathcal{L} &=& -i(\mathbb{I}\otimes H - H^T\otimes \mathbb{I}) \nonumber \\
    &&+ \sum_{j}\left(L^*_j\otimes L_j - \frac{1}{2}(\mathbb{I}\otimes L^{\dagger}_j L_j + (L_j^{\dagger}L_j)^T \otimes \mathbb{I})\right)\,,
    \label{liouvillianotimes}
\end{eqnarray} where we have redefined $\sqrt{\gamma_j}L_j \rightarrow L_j$. Note that we have used the property $ABC = (C^T\otimes A)\bm{B}$ where $A$, $B$, and $C$ are matrices and $\bm{B}$ is the vectorized form of $B$ in column-major form. Next we obtain the matrices $U'$ and $V$ by calculating the singular value decomposition $\mathcal{L} = U' \Sigma V^{\dagger}$. The columns of $U'$ and $V$ corresponding to a singular value equal to zero are denoted by $\bm{\mathcal{U}}'_k$ and $\bm{\mathcal{V}}_k$, respectively, where $k=1,2,...,n$, and $n$ is the number of singular values equal to zero. Note that $\bm{\mathcal{U}}'_k$ and $\bm{\mathcal{V}}_k$ span the null space of $\mathcal{L}^{\dagger}$ and $\mathcal{L}$, respectively. Since $V$ and $U'$ are unitary matrices, the sets $\bm{\mathcal{U}}'_k$ and $\bm{\mathcal{V}}_k$ are orthonormal; however, to use Eq.~\eqref{eq:Albert} we require these sets to be biorthogonal, that is $(\bm{\mathcal{U}}'_k)^\dagger \bm{\mathcal{V}}_{j} = \delta_{kj}$, while keeping $\bm{\mathcal{V}}_k$ to be orthonormal. 

To enforce biorthogonality, we define $V_{\rm null}$ to be the matrix with columns given by $\bm{\mathcal{V}}_k$, and similarly $U'_{\rm null}$ is the matrix with columns given by $\bm{\mathcal{U}}'_k$. We define the overlap matrix $M_O = V^{\dagger}_{\rm null} U'_{\rm null}$, and define a new matrix $U_{\rm null} = U'_{\rm null}M_O^{-1}$. In that way, the columns of $U_{\rm null}$, which we denote by $\bm{\mathcal{U}}_k$, span the null space of $\mathcal{L}^{\dagger}$ but are now biorthogonal with the set $\bm{\mathcal{V}}_k$. Note that the set $\bm{\mathcal{U}}_k$ is, in general, not orthonormal. Finally, once the sets $\bm{\mathcal{U}}_k$ and $\bm{\mathcal{V}}_k$ are obtained, one can use Eq.~\eqref{eq:Albert} to compute the steady states for any initial state $\bm{\rho}_0$.
\crr{
\section{Quantum trajectories and connection to classical attractors}\label{app_C}

As we show in the main text, generically, the steady state of the Lindbladian evolution follows the structure $\bm{\rho}_{\rm{SS}}  = \sum_{j=1}^n c_j \,\bm{\rho}_{{\rm SS},j}$, where $\bm{\rho}_{{\rm SS},j}$ are a set of vectors. In the case where these vectors represent valid density matrices, we can draw some connections to classical attractors. If we consider the formalism of quantum trajectories~\cite{Daley04032014}, we can explore if every quantum trajectory will always end in one of the states $\bm{\rho}_{{\rm SS},j}$. For the case of an ensemble of $N$ qubits evolving under collective decay with jump operator $L_1=S_-$ and rate $\gamma$ (Fig.~\ref{Fig3TwoEnsemble}), the steady state is given by Eq.~\eqref{ssUnb} if the system is initially prepared in the state $\vert \psi_{\rm dif} \rangle$. In that case, the vectors $\bm{\rho}_{{\rm SS},j}$ correspond to the Dicke states $\ket{S,-S}$ with $S=0,1,..,N/2$. All these states are dark states of the jump operator, namely $S_- \ket{S,-S} = 0$, and since there is no coupling between these dark states, each quantum trajectory will indeed converge to one of the Dicke states. As an example, we show quantum trajectories for $N=4$ in Fig.~\ref{FigTraj}, with each trajectory converging to one of the states $\ket{0,0}$, $\ket{1,-1}$, or $\ket{2,-2}$, for long times.

\begin{figure}[t!]
\includegraphics[width=0.45\textwidth]{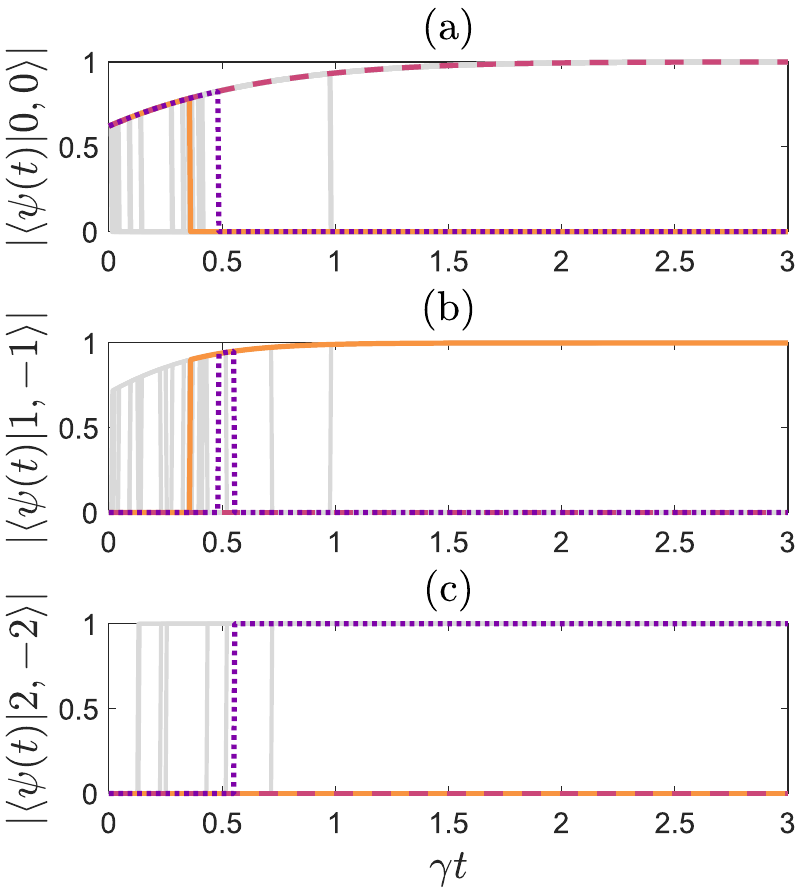}
\caption{\crr{\textbf{Collective decay quantum trajectories:} Quantum trajectories for a system of $N=4$ qubits decaying under collective dissipation with jump operator $L_1=S_-$ and rate $\gamma$. Each trajectory is initiated in the $\vert \psi_{\rm dif} \rangle$ state. Each panel shows the evolution of the overlap with the Dicke state \textbf{(a)} $\ket{0,0}$, \textbf{(b)} $\ket{1,-1}$, and \textbf{(c)} $\ket{2,-2}$. For long times, all individual trajectories converge to one of these states. The same 20 trajectories are shown in each panel, with three trajectories highlighted in different color and line styles to exemplify a trajectory converging to each of the three Dicke states.}}
\label{FigTraj}
\end{figure}

We note that although cases of this kind follow an attractor behavior in the sense that each trajectory converges to one of the states $\bm{\rho}_{{\rm SS},j}$, this is still considerably different from the classical case. Instead of having a given initial state always converge to the same attractor, in the quantum case, the same initial state can converge to different attractors dictated by the probabilities $c_j$, in this case given by $p(S)$. 

Now, for a general case, the condition of the vectors $\bm{\rho}_{{\rm SS},j}$ being valid density matrices is not sufficient to ensure that the trajectories will display this behavior, where they always converge to one of the attractors. A simple case illustrating this is a single qubit dephasing with the jump operator $\sigma_z$. The steady state is described by $\rho_{\rm SS} = p_0 \ket{0}\bra{0} + (1-p_0) \ket{1}\bra{1}$, where $p_0$ is the initial population in state $\ket{0}$. If we initially prepare the system in state $\ket{x_+} = \frac{1}{\sqrt{2}}(\ket{0}+\ket{1})$, the jumps induced by the operator $\sigma_z$ only change the phase of the superposition, but the system never converges to either $\ket{1}$ or $\ket{0}$ for a given trajectory. 

}
\section{Calculation of the Quantum Fisher information  for differential sensing protocol}\label{app_D}

We consider that the signal $\phi$ is encoded by the generator $G=S_z^- = S_z^A -S_z^B$. For a pure state $\ket{\psi}$, the quantum Fisher information is given by:

\begin{equation}
F^Q_{\ket{\psi}} = 4\left(\bra{\psi}(S_z^-)^2 \ket{\psi} - \bra{\psi}S_z^- \ket{\psi}^2\right)\,.
\label{eq:QFI1}
\end{equation}

We note that $F^Q_{\ket{\psi}}$ is proportional to the variance of the generator with respect to the state $\ket{\psi}$. For the cases studied here, the steady state $\rho_{\rm SS}$ will generally be a mixed state with spectral decomposition $\rho_{\rm SS} = \sum_{j=1}^r p_j \ket{\psi_j} \bra{\psi_j}$, where $r$ is the rank. In the case where $\rho_{\rm SS}$ is not full rank, the QFI can be computed as~\cite{Liu_2014}:

\begin{equation}
F^Q_{\rho_{\rm SS}} = \sum_j^r p_j F^Q_{\ket{\psi_j}} - \sum_{j\neq k}^r \frac{8p_jp_k}{p_j+p_k} \vert \langle \psi_j \vert S_z^- \ket{\psi_k}\vert^2\,.
\label{eq:QFI2}
\end{equation}

We use Eq.~\eqref{eq:QFI2} to compute the values of $F^Q_{\rho_{\rm SS}}$ reported in Figs.~\ref{Fig3TwoEnsemble}, ~\ref{Fig4TwoEnsemble}, and~\ref{Fig5}. We note that, as we show below, for all steady states considered here, the second term of the expression above always vanishes, and the QFI reduces to the convex sum over the pure states.

\section{Analytical expression of  the  Fisher information for differential sensing protocols} \label{app_E}

The Clebsch-Gordan coefficients $C_S = \langle S,0 \vert N/4, +N/4, N/4, -N/4\rangle $ can be written explicitly as:

\begin{equation}
C_S = \sqrt{2S+1} \sqrt{\frac{(N/2 !)^2}{(N/2-S)! (N/2+1+S)!} }\,.
\label{CGexp}
\end{equation}

As shown in~\cite{Raphael2025}, we can explicitly compute the QFI of the Dicke states $\ket{S,M}$ using Eq.~\eqref{eq:QFI1}. First, we note that in the case where $N_A=N_B=N/2$, the operator $S_z^-$ can be written in the Dicke basis as:

\begin{eqnarray}
    S_z^-=\sum_{S=1}^{N/2}\sum_{M=-S+1}^{S} &\sqrt{\tfrac{\left(S^2 - M^2\right)\left(\left(N / 2 + 1\right)^2-S^2\right)}{4S^2-1}}\notag\\
    &\times\big( \ket{S-1,M} \langle S, M|+{\rm h.c.}\big)\,.
    \label{encoder}
\end{eqnarray}

Using this result we find:

\begin{eqnarray}
     F_{\ket{S,M}}^{\rm Q} &=& 4\Big(\langle S,M \vert S_z^- S_z^- \vert J,S\rangle - \langle S,M\vert S_z^- \vert S,M \rangle^2 \Big) \notag \\
    =&& \frac{12 M^2 + 8 S (1 + S) (S + S^2 - M^2-1) }{3 - 4 S (1 + S)} \notag \\
    &&+ \frac{(1 - 2 S (1 + S) + 2 M^2)}{3 - 4 S (1 + S)}(N^2 + 4 N). 
    \label{QFIDicke}
\end{eqnarray}

Now, to compute the QFI of the density matrix $\rho_{{\rm SS},L_1}$, we need to use Eq.~\eqref{eq:QFI2}. As seen in Eq.~\eqref{encoder}, the operator $S_z^-$ couples only Dicke states with the same $M$ and total spin $S$ differing by $\pm 1$. Since $\rho_{{\rm SS},L_1}$ is a mixture of states with different $M$ the second term on Eq.~\eqref{eq:QFI2} vanishes and the QFI of this state can be computed by the convex sum $F^Q_{\rho_{{\rm SS},L_1}} = \sum_{S=0}^{N/2} p(S) F^Q_{\ket{S,-S}}$. Given that we have analytical expressions for $p(S)$ in Eq.~\eqref{CGexp} and $F^Q_{\ket{S,-S}}$ in Eq.~\eqref{QFIDicke}, we can numerically evaluate $F^Q_{\rho_{{\rm SS},L_1}}$ for any $N$. We use this to obtain the dashed orange line in Fig.~\ref{Fig4TwoEnsemble}(d).

In a similar manner, we can explicitly compute the QFI of the states $\rho_{{\rm B},S}$. We note that these states are a mixture of Dicke states with the same value of $S$, according to Eq.~\eqref{encoder}, the operator $S_z^-$ does not couple these states, and the QFI is simply given by the first term in Eq.~\eqref{eq:QFI2}. Then, the QFI is given by $F^Q_{\rho_{{\rm B},S}}=\frac{1}{2S+1}\sum_{M=-S}^S F^{Q}_{\ket{S,M}}$. The sum can be simplified to obtain the closed expression $F^Q_{\rho_{{\rm B},S}} = \frac{N^2+4N}{3}-\frac{4}{3}S(S+1)$, as we report in Fig.~\ref{Fig4TwoEnsemble}(b).

Finally, we can compute the QFI of the steady state generated by the protocol proposed in the main text [see Fig.~\ref{Fig4TwoEnsemble}(c)]. Since in each run of the protocol a given state $\ket{S,-S}$ is obtained with probability $p(S)$ after the evolution under $L_1 = S_-$, the QFI is given by $F^Q_{\rm  pro} = \sum_{S=0}^{N/2} p(S) F^Q_{\rho_{{\rm B},S}}$. Using the analytical expressions derived above, we obtain $F^Q_{\rm pro} = \frac{N^2 +2 N}{3}$, which shows the Heisenberg scaling of the states generated with the protocol [see solid line in Fig.~\ref{Fig4TwoEnsemble}(d)]. We note that the results derived in this Appendix for the QFI of different states rely on the condition $N_A=N_B=N/2$. If we have an imbalance $\eta$, it is hard, in general, to derive analytical expressions and one has to approach the task numerically, see Fig.~\ref{Fig5}, for example.

\end{document}


\title{Supplemental Material: Generating Entangled Steady States in Multistable Open Quantum Systems via Initial State Control}

\author{Diego Fallas Padilla}
\email{difa1788@colorado.edu}
\affiliation{JILA, NIST, and Department of Physics, University of Colorado, Boulder, CO, 80309, USA}
\affiliation{Center for Theory of Quantum Matter, University of Colorado, Boulder, CO 80309, USA}
\author{Raphael Kaubruegger}
\affiliation{JILA, NIST, and Department of Physics, University of Colorado, Boulder, CO, 80309, USA}
\affiliation{Center for Theory of Quantum Matter, University of Colorado, Boulder, CO 80309, USA}
\author{Adrianna Gillman}
\affiliation{Department of Applied Mathematics, University of Colorado, 526 UCB, Boulder, CO, 80309, USA}
\author{Stephen Becker}
\affiliation{Department of Applied Mathematics, University of Colorado, 526 UCB, Boulder, CO, 80309, USA}
\author{Ana Maria Rey}
\affiliation{JILA, NIST, and Department of Physics, University of Colorado, Boulder, CO, 80309, USA}
\affiliation{Center for Theory of Quantum Matter, University of Colorado, Boulder, CO 80309, USA}
\date{\today }
\maketitle

\par In this Supplementary Material, we give more details about the derivation of Eq.~(3) of the main text. Moreover, we provide two additional examples where we illustrate the use of the formalism presented in the main text. The first one is an extension of the two-qubit example provided in the main text, but now with a non-zero Hamiltonian. The second describes multi-level atoms undergoing collective decay, and we use it to showcase the numerical efficiency of our expressions versus integrating the time dynamics for long times.

\section{Derivation of the main results: Additional details}

The starting point is Eq.~(A6) of Appendix A in the main text:

\begin{equation}
\bm{\rho}_{\rm SS} = V \left(\lim_{s \rightarrow 0}s\left(s U^{\dagger}V - \Sigma \right)^{-1}\right) U^{\dagger}\bm{\rho}_0.
\label{eqSVD}
\end{equation}

We now focus on computing the limit $\lim_{s \rightarrow 0}s\left(s U^{\dagger}V - \Sigma \right)^{-1}$. Let us start by rewriting the different matrices in block form:

\begin{eqnarray}
\Sigma = \begin{pmatrix}
\Sigma_{11} & 0 \\
0 & 0
\end{pmatrix}, \quad U^{\dagger} V = A = \begin{pmatrix}
A_{11}&A_{12} \\
A_{21}&A_{22}\,
\end{pmatrix},
\label{block}
\end{eqnarray}
where $\Sigma_{11}$ is a diagonal matrix with non-zero diagonal elements, consequently, $\Sigma_{11}$ is invertible. We are interested in computing the inverse $(sA-\Sigma)^{-1}$. To do so, we make use of the Schur complement~\cite{zhang2006schur} to write:

\begin{eqnarray}
(sA-\Sigma)^{-1} = \begin{pmatrix}
\alpha_{11} & \alpha_{12} \\
\alpha_{21} & \alpha_{22}\,
\end{pmatrix},
\end{eqnarray}
where each block is defined as follows:

\begin{eqnarray}
&&\alpha_{11} = (-\Sigma_{11}+s A_{11})^{-1} + s^2 (-\Sigma_{11}+s A_{11})^{-1} A_{12} (\zeta(s))^{-1}A_{21}(-\Sigma_{11}+s A_{11})^{-1}, \nonumber \\
&&\alpha_{12} = -s (-\Sigma_{11}+s A_{11})^{-1} A_{12} (\zeta(s))^{-1},\, \alpha_{21} = -s (\zeta(s))^{-1}  A_{21} (-\Sigma_{11}+s A_{11})^{-1}, \nonumber \\
&&\alpha_{22} = (\zeta(s))^{-1},\, \zeta(s) = sA_{22} - s^2A_{21} (-\Sigma_{11}+s A_{11})^{-1}A_{21}\,.
\label{alphas}
\end{eqnarray}

Here $\zeta(s)$ is the Schur complement of the top left block of the matrix $sA -\Sigma$. In a couple of the terms above, we need to compute the inverse of a matrix that is perturbed, namely, $(B_1 +s B_2)^{-1}$, with $s \rightarrow 0$. To compute this, we can expand the inverse using a Neumann series:

\begin{equation}
(B_1 - B_2)^{-1} = \sum_{k=0}^\infty ((B_1)^{-1}B_2)^k (B_1)^{-1},
\end{equation}
which is valid if the spectral radius of $((B_1)^{-1}B_2)$ is less than one. Applying this formula to the case $(B_1 +s B_2)^{-1}$, we obtain:

\begin{equation}
(B_1 -(-s B_2))^{-1} = (B_1)^{-1} - s (B_1)^{-1} B_2 (B_1)^{-1} + \mathcal{O}(s^2). 
\end{equation}

Using this equation, we find:

\begin{eqnarray}
&&(-\Sigma_{11}+s A_{11})^{-1} \approx -\Sigma_{11}^{-1}- s \Sigma_{11}^{-1}  A_{11} \Sigma_{11}^{-1}, \nonumber \\
&& (\zeta(s))^{-1} \approx \frac{1}{s} A_{22}^{-1} -  A_{22}^{-1} (A_{21}\Sigma_{11}^{-1}A_{21})  A_{22}^{-1}. 
\label{pert}
\end{eqnarray}

Substituting the results in Eq.~\eqref{pert} into Eq.~\eqref{alphas}, and taking the limit $\lim_{s \rightarrow 0} s \alpha_{ij}$ we find that $s \alpha_{11}$, $s \alpha_{12}$, and $s \alpha_{21}$ are all of order 1 in $s$ at the lowest order, while $s\alpha_{22} = A_{22}^{-1} + \mathcal{O}(s)$. Then we can rewrite Eq.~\eqref{eqSVD} as:

\begin{equation}
\bm{\rho}_{\rm SS} = V \begin{pmatrix}
    0&0\\
    0&A_{22}^{-1}
\end{pmatrix} U^{\dagger}\bm{\rho}_0.
\label{eqSVDsimp}
\end{equation}

Note that according to $A = U^\dagger V$ and the block definitions of both $A$ and $\Sigma$ in Eq.~\eqref{block}, $A_{22} = U_{\rm null}^\dagger V_{\rm null}$, where $U_{\rm null}$ is composed of the columns of $U$ corresponding to a singular value of zero, similarly, $V_{\rm null}$ is defined by the columns of $V$ corresponding to zero singular value. In the main text, we denote the columns of $U_{\rm null}$ by $\bm{\mathcal{U}}_j$ and the ones of $V_{\rm null}$ by $\bm{\mathcal{V}}_j$. As described in Appendix B of the manuscript, we can construct these two sets of columns to be biorthogonal, that is $\bm{\mathcal{U}}_j^\dagger \bm{\mathcal{V}}_k = \delta_{jk}$. Consequently, $A_{22} = U_{\rm null}^\dagger V_{\rm null} = \mathbb{I}_n $, where $n$ is the number of zero singular values and $\mathbb{I}_n$ is the identity. We can then rewrite Eq.~\eqref{eqSVDsimp} as:

\begin{equation}
\bm{\rho}_{\rm SS} = V \begin{pmatrix}
    0&0\\
    0&\mathbb{I}_n
\end{pmatrix} U^{\dagger}\bm{\rho}_0.
\label{eqSVDsimp2}
\end{equation}

We note that this expression is very similar to Eq.~(A10) in Appendix A of the manuscript. Following very similar steps to the ones described in that Appendix, we can rewrite this expression as:

\begin{equation}
\bm{\rho}_{\rm SS} = \sum_{j=1}^n  \bm{\mathcal{U}}^{\dagger}_j  \bm{\rho}_0  \bm{\mathcal{V}}_j\,,
\label{eqSVD3}
\end{equation} which is Eq.~(3) of the main text.

\section{Two qubits with collective dissipation: The effect of a drive}

\begin{figure}[t!]
\includegraphics[width=0.7\textwidth]{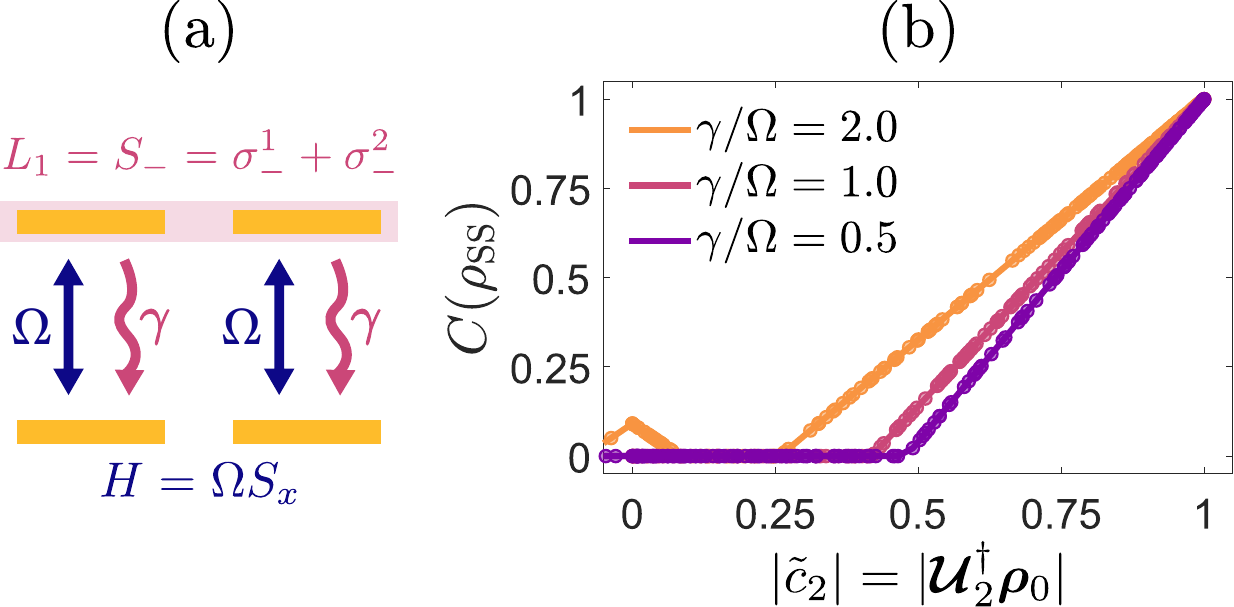}
\caption{\textbf{Two qubits under collective dissipation and drive:} \textbf{(a)} Two qubits undergo dynamics generated by the collective dissipation jump operator $L_1=S_-=\sigma_-^1+\sigma_-^2$ with rate $\gamma$, and a Hamiltonian drive given by $H=\Omega S_x = \Omega (\sigma_x^1+\sigma_x^2)$. \textbf{(b)} Concurrence in the steady state as a function of the overlap $\tilde{c}_2 = \bm{\mathcal{U}}_{2}^\dagger \bm{\rho}_0$ for the setup in (a). Data represents 300 initial states $\bm{\rho}_0$ sampled at random. The steady state is computed using Eq.~(5) in the main text with $\epsilon = 0.0001$ (solid line), and with Eq.~(3) in the main text (markers).}
\label{Fig1supp}
\end{figure}

We consider a system of two qubits evolving under collective dissipation with jump operator $L_1 = S_-$ and rate $\gamma$. Unlike the study in the main text, we now consider the presence of a Hamiltonian drive of the form $H = \Omega S_x$ [see Fig.~\ref{Fig1supp}(a)]. Again, we label the two states of each qubit as $\ket{0}$ and $\ket{1}$, and describe the full Hilbert space in the product basis $\{\ket{1} = \ket{1}\otimes \ket{1}, \ket{2}=\ket{1}\otimes \ket{0}, \ket{3} = \ket{0}\otimes \ket{1}, \ket{4}=\ket{0}\otimes \ket{0}\}$. As we did for the case with $H=0$ in the main text, we use Eq.~\eqref{eqSVD3} to explore the form of the steady state. With the presence of the Hamiltonian term, we find that the null space of $\mathcal{L}$ contains only two elements (different from $n=4$ when $H=0$). $\bm{\mathcal{V}}_2$ is still the singlet state, namely, $\mathcal{V}_2 = \ket{\phi_-}\bra{\phi_-}$ with $\ket{\phi_{\pm}} = \frac{1}{\sqrt{2}}(\ket{2}\pm \ket{3})$. On the other hand, the vector $\bm{\mathcal{V}}_1$ is a mixture of pure states, with each pure state being a linear superposition of the states $\ket{1}$, $\ket{4}$, and $\ket{\phi_+}$. The specific form of $\bm{\mathcal{V}}_1$ is highly dependent on the ratio $\gamma /\Omega$. We note that the symmetries of the system are still present in the vectors spanning the steady state, with $\bm{\mathcal{V}}_1$ involving only states with $S=1$, while $\bm{\mathcal{V}}_2$ corresponds to the $S=0$ state.

We find that we can set $\bm{\mathcal{V}}_2 = \bm{\mathcal{U}_2}$, but $\bm{\mathcal{U}}_1 \neq \bm{\mathcal{V}}_1$ (see the discussion in the main text on why it is important that these two vectors are different), with its form depending on $\gamma/\Omega$. Since $\bm{\mathcal{V}}_2 = \bm{\mathcal{U}_2}$, we can still consider the overlap with the singlet as the parameter we modify to study the entanglement of the steady state; however, note that now the entanglement content of $\bm{\mathcal{V}}_1$ varies as a function of $\gamma/\Omega$, then, in general, we expect that for some values of this ratio the relation between the entanglement in the steady state and the overlap with $\bm{\mathcal{V}}_2$ will be non monotonic. 

In Fig.~\ref{Fig1supp}(b), we show the concurrence as a function of the overlap with the singlet for different values of the ratio $\gamma/\Omega$. We note that for $\gamma/\Omega > 1$ there is a possibility to have non-zero entanglement for small values of the overlap $\bm{\mathcal{U}}_{2}^\dagger \bm{\rho_0}$, as discussed above, this is expected as $\bm{\mathcal{V}}_1$ can have non-zero entanglement content depending on the ratio of these parameters. We note that both methods in Eqs. (3) and (5) of the main text agree with each other, even in the presence of a non-zero Hamiltonian; however, we should note that the presence of the Hamiltonian makes the structure of the vectors $\bm{\mathcal{V}}_j$ more complex. In general, for more complex Hamiltonians, obtaining analytical expressions for the vectors $\bm{\mathcal{V}}_j$ or $\bm{\rho}_{{\rm SS},j}$ might not be accessible; nonetheless, all the methods described in the main text can still be implemented numerically, with some of them being preferred over others depending on the setup, as discussed in the main text.

\section{Dark states of multi-level atoms}

\begin{figure}[t!]
\includegraphics[width=0.7\textwidth]{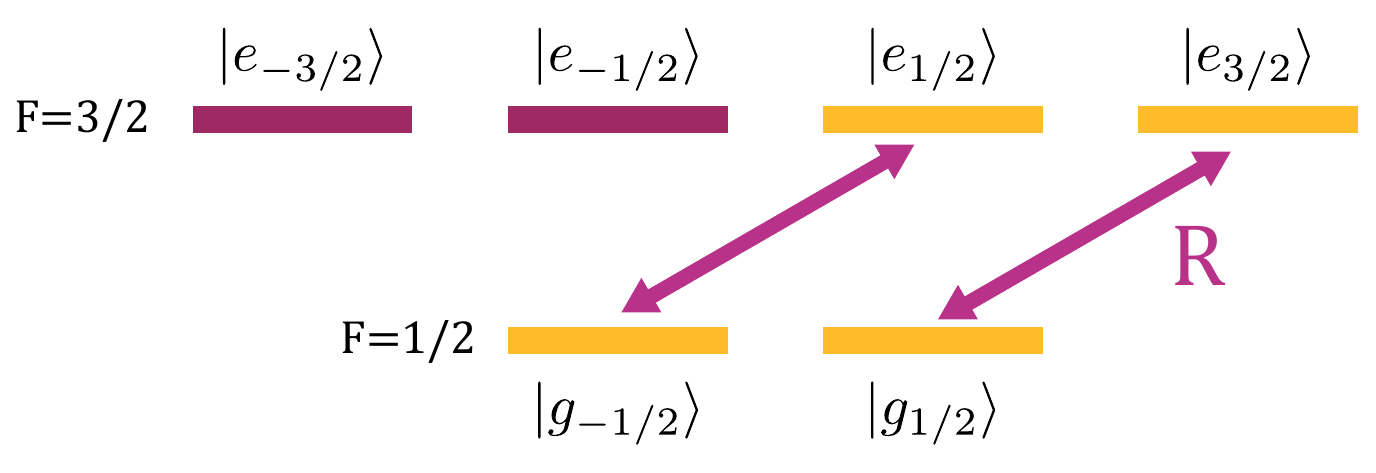}
\caption{\textbf{Multi-level atoms coupled by circularly polarized light:} We consider atoms composed of six hyperfine levels, two ground states with $F=1/2$ and four excited states corresponding to $F=3/2$. The states are labeled according to the value of $m_F$. If initially the system has population only on the ground states, coupling to the excited state manifold using right-handed circularly polarized light [with dipole operator $R$ described in Eq.~\eqref{ROp}] reduces the dynamics to the four levels colored in yellow.}
\label{Fig2supp}
\end{figure}

The examples provided so far consist of systems of two-level atoms (qubits) evolving under collective dynamics. However, the methods presented here are general and can be applied to more complex systems. The richness of the possible steady states that can be found in an atom ensemble increases when multi-level atoms are considered. Here we consider the setup described in Section V of Ref.~\cite{pineiroorioli2022}. The system is composed of $N$ atoms, each one containing two hyperfine ground states with $F=F_g = 1/2$ and four excited states corresponding to $F=F_e = 3/2$ as depicted in Fig.~\ref{Fig2supp}. We label the ground states as $\ket{g_{m_F}}$ and the excited states as $\ket{e_{m_F}}$.

In multi-level atoms, different internal levels can be coupled through light with different polarizations. Here, we will consider right-handed circularly polarized light that couples states $\ket{g_{-1/2}}$ and $\ket{g_{1/2}}$ to states $\ket{e_{1/2}}$ and $\ket{e_{3/2}}$, respectively. If the system is initialized with all atoms in the ground states, only four out of the six internal levels get involved in the dynamics as shown in Fig.~\ref{Fig2supp}. We can understand this coupling through the collective dipole operator $R$ defined as:

\begin{equation}
R = \frac{1}{2} (R_- + R_+),\, \,R_- = (R_+)^\dagger =  \sigma_{g_{1/2},e_{3/2}} + \frac{1}{\sqrt{3}} \sigma_{g_{-1/2},e_{1/2}},
\label{ROp}
\end{equation}
where $\sigma_{a,b}$ are collective operators defined by $\sigma_{a,b} = \sum_{j=1}^N \vert a \rangle_j \langle b \vert_j$, with the sum on index $j$ being over all atoms. The factor of $\frac{1}{\sqrt{3}}$ is related to the Clebsch-Gordon coefficient for the transition between those two internal states, namely, $\langle F_g,m,1,+1\vert F_e,m+1\rangle = 1/\sqrt{3}$ for $F_g=1/2$, $F_e=3/2$, and $m=-1/2$. In Ref.~\cite{pineiroorioli2022}, the authors explored how the collective decay under the jump operator $L_1 = R_-$ can lead to interesting steady states depending on the initial state. The dynamics studied there can be described as follows: first, the atoms are prepared in a product state where every atom is in the singlet state $\ket{\psi_{\rm singlet}}=\frac{1}{\sqrt{2}}(\vert g_{1/2}\rangle - \vert g_{-1/2} \rangle)$. This product state is denoted by $\ket{\psi_{\rm prod}} = \ket{\psi_{\rm singlet}}^{\otimes N}$. Then, the atoms are driven by the Hamiltonian $H = \Omega R$ for a certain time $\tau$. Finally, the system decays collectively under the jump operator $L_1=R_-$, with rate $\Gamma$, until it reaches a steady state. For certain values of $\theta_0=\Omega \tau$, the steady state can contain some population in the excited manifold. 

\begin{figure}[t!]
\includegraphics[width=0.96\textwidth]{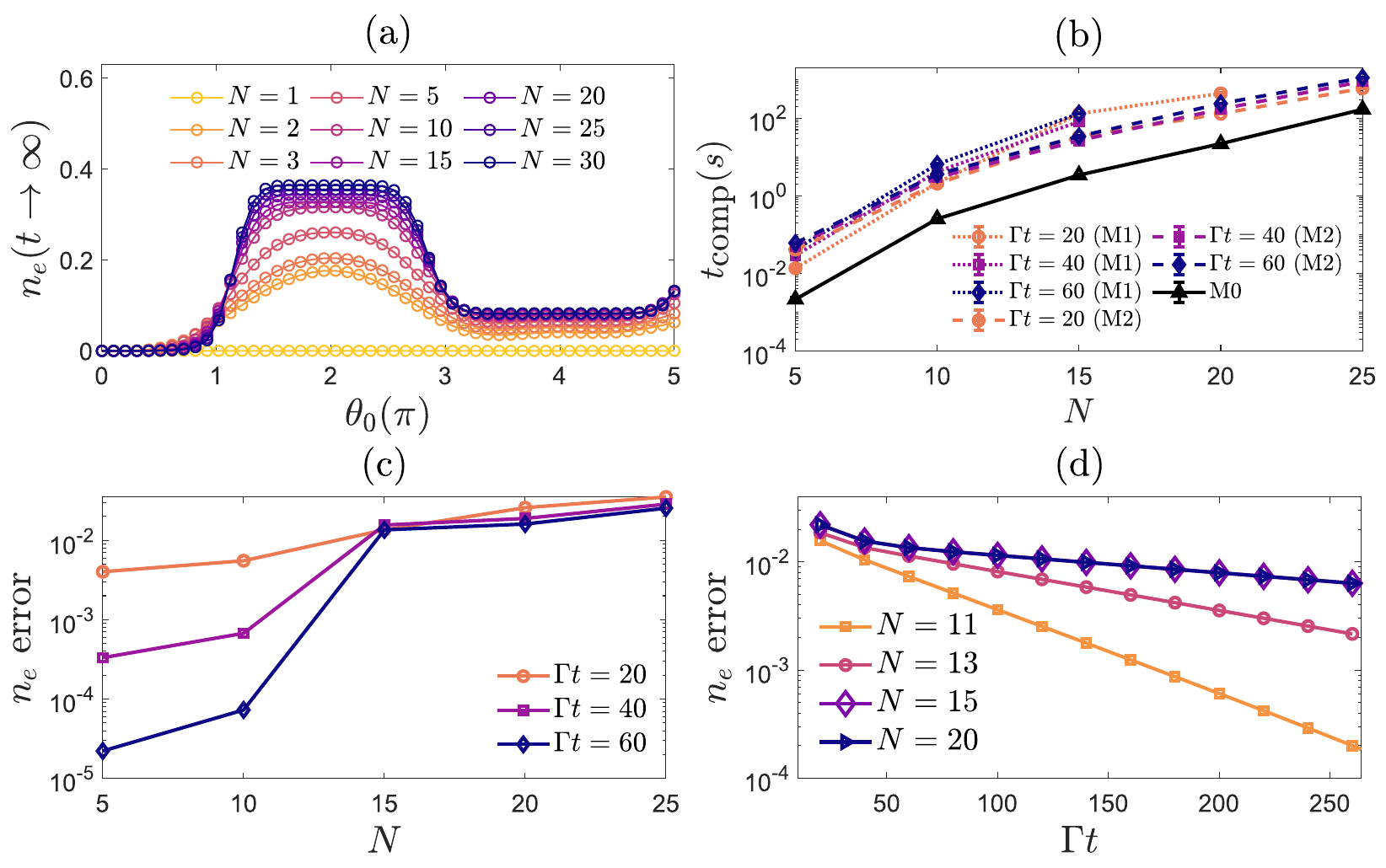}
\caption{\textbf{Multi-level atom steady states:} {\bf(a)} Excited state population in the steady state as a function of $\theta_0=\Omega \tau$. Data was obtained by solving Eq.~(5) in the main text. Different colors denote different numbers of atoms. This should be compared to Fig.~5(c) in Ref.~\cite{pineiroorioli2022}. 
{\bf (b)} Computational time it takes to compute the steady state for our method (M0), a Runge-Kutta ODE solver (M1), and a matrix exponentiation method (M2) as a function of N for the case $\theta_0 = 2\pi$. For methods M1 (dotted) and M2 (dashed), different integration times $\Gamma t$ are signaled by different colors. Each data point is the average of 50 computations of the same numerical task, with the error bars showing a standard deviation at least one order of magnitude smaller than the average.
{\bf(c)} Absolute error of the excited state population $n_e$ as a function of $N$ for $\theta_0 = 2\pi$. The error represents the difference between the results obtained with the M0 and M2 methods. Different colors and markers represent different integration times $\Gamma t$ for M2.
{\bf(d)} Absolute error of the excited state population as a function of $\Gamma t$ for $\theta_0 = 2\pi$. The error is defined as in (c). Different colors and markers correspond to different $N$. For all panels, we consider $\epsilon = 0.0001$ when using Eq.~(5) in the main text.
}
\label{Fig3}
\end{figure}

Here, we confirm whether our formalism, specifically Eq.~(5) in the main text (we refer to this method as M0 in Fig.~\ref{Fig3}), can be used to reproduce these results. We quantify the population in the excited state manifold through the expectation value of the operator $n_e = \frac{1}{N}(\sigma_{e_{1/2},e_{1.2}}+\sigma_{e_{3/2},e_{3/2}})$. Here, we evolve the initial state $\ket{\psi_{\rm prod}}$ under the Hamiltonian $H=\Omega R$ for different values of $\theta_0$ using conventional ED techniques, the resulting state corresponds to $\ket{\psi_0}= e^{-iH\tau} \ket{\psi_{\rm prod}}$. Finally, we compute the steady state of the dissipative dynamics with jump operator $L_1=R_-$ for the initial state $\rho_0 = \ket{\psi_0}\bra{\psi_0}$ using Eq.~(5) of the main text. The results are shown in Fig.~\ref{Fig3}(a).

The results obtained using our method (M0) are consistent with the ones reported in Fig.~5(c) in Ref.~\cite{pineiroorioli2022}. Up to $N=10$, the results agree to a very high accuracy. For $N>15$ the results obtained predict a lower excited state population compared the ones in that work. However, through simulation of the time dynamics for very long times, we confirm that the results obtained with our method are accurate, as they converge to the time dynamics results when $\Gamma t \rightarrow \infty$, as reported in Fig.~\ref{Fig3}(d).

In Fig.~\ref{Fig3}(c), we show how, for a fixed $\Gamma t$, the discrepancy between the excited state population obtained through our method and the numerical integration methods (see below) increases as a function of $N$. This means that for this multi-level system, the time it takes the system to reach the steady state increases as a function of $N$, as visible in Fig.~5(b) in Ref.~\cite{pineiroorioli2022}. A longer time to reach the steady state and a larger system size can both negatively affect the computation time elapsed during the time integration, contrary to our method which is, in principle, only affected by the latter. We explore the computational time differences in Fig.~\ref{Fig3}(b).

The integration of the Lindblad equation (Eq.~(1) in the main text), where the Liouvillian $\mathcal{L}$ is represented by a large sparse matrix, can be performed using multiple methods, such as conventional ODE Runge-Kutta methods or more sophisticated sparse-matrix exponentiation routines. Here, we consider the MATLAB implementation of three methods: (i) solving Eq.~(5) utilizing the \texttt{mldivide} routine, (ii) a Runge-Kutta (4,5) method (M1) through the \texttt{ode45} solver \cite{DORMAND198019,MATLABODE}, and (iii) a Krylov subspace sparse-matrix exponentiation method (M2) introduced in the EXPOKIT package \cite{EXPOKIT}. As reported in Fig.~\ref{Fig3}(b), computing the steady state with our method is, at least, one order of magnitude faster than the M1 and M2 methods for all $N$  considered here. Moreover, both M1 and M2 get slower as a function of $\Gamma t$. Reducing $\Gamma t$ to reduce computational time, would compromise the precision of the steady-state observable values [panels (c) and (d)]. 

\subsection{Computational details}

The Krylov subspace exponentiation method for sparse matrices (M2), can provide the specific solution for $\Gamma t$ without the need of storing intermediate states. Nonetheless, Krylov subspace methods solving the equation $\bm{\rho}(t) = \exp(\mathcal{L}t)\bm{\rho}_0$ become considerably slower as $\Gamma t$ grows. Then, one can consider a grid of intermediate states separated by time intervals $\Delta t_i$, such that each evolution $\bm{\rho}(t_{i+1}) = \exp(\mathcal{L}\Delta t_i)\bm{\rho}(t_i)$ is faster, and then optimize the number and magnitude of intermediate steps. Here, we did not consider any intermediate steps.

All calculations for the computational time benchmark were performed in the JILA cluster. For each data point in Fig.~\ref{Fig3}(b), the numerical calculations were performed in a designated isolated node (used only for this numerical task). This node contains 24 cores and 180 GB of memory with CPU model Intel Xeon Silver 4214 @ 2.2GHz.

\bibliographystyle{apsrev4-2}
\bibliography{refs}{}